\documentclass[twocolumn,amsmath,amssymb,showpacs]{revtex4}

\usepackage{amsmath}
\usepackage{graphicx}

\begin{document}
\title{Graphene's morphology and electronic properties from discrete differential geometry}
\author{Alejandro A. Pacheco Sanjuan,$^{1}$  Zhengfei Wang,$^{2}$\\  Hamed Pour Imani,$^{3}$ Mihajlo Vanevi\'c,$^4$ and Salvador Barraza-Lopez$^3$}
\affiliation{1. Departamento de Ingenier{\'\i}a Mec\'anica. Universidad del Norte. Km.~5 V{\'\i}a Puerto Colombia. Barranquilla, Colombia\\
2. Department of Materials Science. University of Utah.  Salt Lake City, UT 84112, USA\\
3. Department of Physics. University of Arkansas. Fayetteville, AR 72701, USA\\
4. Department of Physics. University of Belgrade. 11158 Belgrade, Serbia}

\begin{abstract}
The geometry of two-dimensional crystalline membranes dictates their mechanical, electronic and chemical properties. The local
geometry of a surface is determined from the two invariants of the metric and the curvature tensors. Here we discuss those invariants directly from atomic positions
in terms of angles, areas, vertex and normal vectors from carbon atoms on the graphene lattice, for arbitrary elastic regimes and atomic conformations, and without
recourse to an effective continuum model. The geometrical analysis of graphene membranes under mechanical load is complemented with a study of the local density of states
(LDOS), discrete induced gauge potentials, velocity renormalization, and non-trivial electronic effects originating from the scalar deformation
potential. The asymmetric LDOS is related to sublattice-specific deformation potential differences, giving rise to the pseudomagnetic field.
The results here enable the study of geometrical, mechanical and electronic properties for arbitrarily-shaped graphene membranes in
experimentally-relevant regimes without recourse to differential geometry and continuum elasticity.
\end{abstract}
\date{\today}
\maketitle

\noindent{\em Introduction.-} Graphene \cite{Wallace,RMP} belongs to a family of atom-thin elastic membranes \cite{Novo} that conform to harder
surfaces (e.g., \cite{PRLMorozov2006}), develop ripples when
freestanding \cite{Nature2007,Biro,Shenoy,Liu,HWang,newprl,us,usSSC,Fasolino1,Zakharenko,Katsnelson2,PRB2010,sanjose}, and can be deformed into
arbitrary elastic regimes  \cite{Hone1}, leading to a remarkable electronic
response \cite{Ando2002,Pereira1,GuineaNatPhys2010,Vozmediano,deJuanPRB,Dejuan2011,deJuanPRL2012,Kitt2013,Peeters3,Peeters4,Naumis1,Manes,Hone1,Crommie,usold,stroscio}.
In general, the local geometry of a two-dimensional (2D) surface is determined by four invariants of its metric ($g$) and curvature ($k$), that indicate
 how much it stretches and curves with respect to a reference non-deformed shape.
Suitable choices are the determinant and the trace of $g$, the Gauss curvature $K\equiv\det(k)/\det(g)$, and the mean
curvature $H\equiv\text{Tr}(k)/2\text{Tr}(g)$ \cite{doCarmo,Lee,M4}.

In the existing literature, graphene's geometry is commonly studied in terms of a continuous displacement field $u_{\alpha}(\xi^1,\xi^2)$. Specifically, on thin-plate continuum
elasticity the strain tensor is
$u_{\alpha\beta}=(\partial_{\alpha}u_{\beta}+\partial_{\beta}u_{\alpha}+ \partial_{\alpha}u_{\beta}\partial_{\beta}u_{\alpha}+ \partial_{\alpha}z\partial_{\beta}z)/2$,
with $z$ an out-of-plane elongation \cite{Ando2002,Pereira1,GuineaNatPhys2010,Vozmediano,deJuanPRB,Dejuan2011,deJuanPRL2012,Kitt2013,Peeters3,Peeters4,Naumis1,Manes}.
 There, differential geometry and mechanics couple as:
\begin{equation}
g_{\alpha\beta}=\delta_{\alpha\beta}+2u_{\alpha\beta},\qquad \text{ }k_{\alpha\beta}=\hat{\mathbf{n}}\cdot \frac{\partial \mathbf{g}_{\alpha}}{\partial \xi^{\beta}},
\end{equation}
where $\mathbf{g}_{\alpha}(\xi^1,\xi^2)$ is a tangent vector field, $\delta_{\alpha\beta}$ is the reference (flat) metric
and $\hat{\mathbf{n}}=\frac{\mathbf{g}_{\xi^1}\times \mathbf{g}_{\xi^2}}{|\mathbf{g}_{\xi^1}\times \mathbf{g}_{\xi^2}|}$ is the local normal.
{\em Strain engineering clearly is a geometrical theory}, and differential geometry is the basis of this formalism
as we know it \cite{Pereira1,GuineaNatPhys2010,Vozmediano,deJuanPRB,Dejuan2011,deJuanPRL2012,Kitt2013,Peeters3,Peeters4,Naumis1,Manes}. However, the
geometrical description given by Eq.~(1) has limitations. Continuum theory usually requires slow-varying, harmonic deformations,
conditions that are violated in realistic situations \cite{Dumitrica2011,Biro}. Peculiarities of how graphene ripples \cite{ChenNatureNano,Biro,M2,M4,Dumitrica2011}, slides and adheres \cite{Biro,Kitt2} may be beyond first-order continuum elasticity.

 This calls for a fundamental study of the geometry of atomistic membranes and their subsequent coupling to electronic degrees of freedom, down to
 unavoidable atomic-scale fluctuations \cite{Fasolino1,Katsnelson2}. Geometry is relevant in addressing spin diffusion in rippled
 graphene \cite{Ando2000,Huertas-Hernando}, in understanding the chemical properties of conformal (non-planar) 2D crystals \cite{ACSNano}, and may even
 herald the strain engineering of 2D crystals with atomistic defects, an area completely unexplored so far.

In this Letter we develop a theoretical framework for strain engineering \cite{us,usSSC,pe2013} based on
discrete geometry, that applies to arbitrarily-shaped graphene without topological defects. Here, Wigner-Seitz unit cells are
the underlying discrete geometrical objects and atomistic information is always
 preserved. The discrete formalism for geometry and the electronic response of Dirac fermions rests on interatomic
 distances without a mediating continuum. The framework here realized is {\em non-perturbative} on the geometry,
 and it can be used to indicate if the sublattice symmetry is preserved in the system at hand
 (this is assumed in the continuum theory \cite{GuineaNatPhys2010}) and to show how the reciprocal space is renormalized by strain \cite{Kitt2013}.
In what follows, we present the tools for geometrical analysis, study the local geometry of rippled
 graphene \cite{Fasolino1}, and discuss the discrete geometry and the electronic properties of graphene under central load.

\begin{figure}[tb]
\begin{center}
\includegraphics[width=0.47\textwidth]{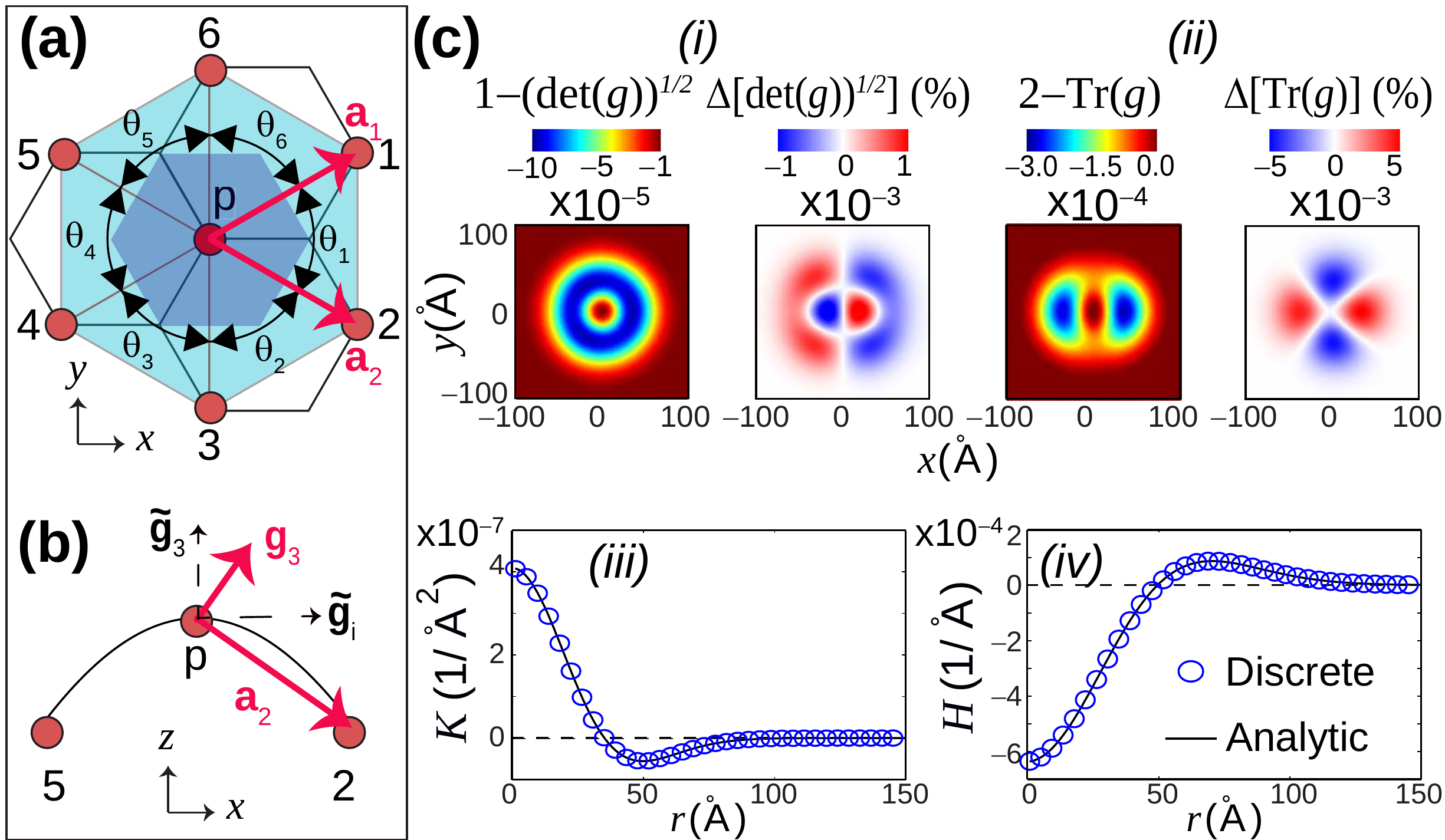}
\end{center}
\caption{(Color online) (a) Top view of polyhedra used to determine the four geometrical invariants from the metric and curvature.
Circles represent atoms on the A-sublattice. Local lattice vectors are $\mathbf{a}_1$ and $\mathbf{a}_2$; $\theta_i$ are internal angles to edges $\mathbf{e}_i$
and $\mathbf{e}_{i+1}$; and the central shaded hexagon is the Voronoi cell. (b) Side view highlights the differences between continuum and discrete vector fields.
  (c) i: $\sqrt{\det(g)}$, ii: $\text{Tr}(g)$, iii: $K$ and iv: $H$ for a smooth gaussian bump where discrete and continuum results coincide. Percent differences
  $\sqrt{\det(\tilde{g})}-\sqrt{\det(g)}$ and $\text{Tr}(\tilde{g})-\text{Tr}(g)$ are also shown.}\label{fig:nodalpoints}
\end{figure}

\noindent{\em The discrete geometry.- }The discrete metric is defined from the local lattice vectors $\mathbf{a}_{\alpha}$ \cite{us,usSSC,SI}
$g_{\alpha\beta}=\mathbf{a}_{\alpha}\cdot \mathbf{a}_{\beta}$ [Fig.~\ref{fig:nodalpoints}(a-b)], and the discrete Gauss curvature ($K_D$) originates
from the {\em angle defect} $\sum_{i=1}^6\theta_i$ \cite{Math1,Math2,chinese}:
\begin{equation}\label{eq:DGB}
K_D=(2\pi-\sum_{i=1}^6\theta_i)/A_p.
\end{equation}
Here $\theta_i$ ($i=1,...,6$) are angles between vertices shown in Fig.~1(a).
The {\em Voronoi tessellation} [dark blue in Fig.~1(a) with an area $A_p$] generalizes the Wigner-Seitz unit cell on conformal 2D geometries.
 (The angle defect adds up to $2\pi$ on a flat surface, making $K_D=0$, as expected.)

 The discrete mean curvature $H_D$ measures relative orientations of edges and normal vectors along a closed path:
\begin{equation}\label{eq:Hdiscrete}
H_{D}=\sum_{i=1}^6 \mathbf{e}_i\times(\boldsymbol{\nu}_{i,i+1}-\boldsymbol{\nu}_{i-1,i})\cdot \hat{\mathbf{n}}/4A_p.
\end{equation}
Here, $\mathbf{v}_i$ is the position of atom $i$ on sublattice $A$, and $\mathbf{e}_i=\mathbf{v}_i-\mathbf{v}_p$ is the {\em edge} between points
$p$ and $i$ (note that $\mathbf{a}_{1(2)}=\mathbf{e}_{1(2)}$). $\boldsymbol{\nu}_{i,i+1}$ is the normal to edges $\mathbf{e}_i$ and $\mathbf{e}_{i+1}$ ($i$
is a cyclic index), and  $\hat{\mathbf{n}}=\frac{\sum_{i=1}^6\boldsymbol{\nu}_{i,i+1}A_i}{\sum_{i=1}^6A_i}$ is the area-weighted normal
with $A_i=|\mathbf{e}_i\times \mathbf{e}_{i+1}|/2$ \cite{Math1}. For the purposes of discrete geometry, the metric and curvatures are formally decoupled objects.

  The discrete metric and curvatures furnish geometry consistent with a crystalline structure, and lead to the faithful characterization of graphene's
  morphology beyond the effective-continuum paradigm, Eq.~(1). This is advantageous when the atomic conformation is known from molecular dynamics
  (e.g, \cite{Fasolino1}) or experiment (e.g., \cite{MolecularGraphene}) because: (i) fitting of the atomic lattice to an effective continuum is not needed
  any more, (ii) the Chemistry of conformal graphene can be addressed from the discrete geometry \cite{ACSNano} and, since atoms are always
  available, (iii) the discrete theory brings new insights and understanding into the physical theory (e.g., non-preservation of sublattice symmetry, the
  form of gauge fields \cite{us}, the creation of mass from strain \cite{usSSC,Manes}). We emphasize that the discrete geometry is accurate regardless of elastic regime, hence
  it can be used to verify whether the conditions for continuum elasticity hold in the problem at hand.

\noindent{\em Continuum geometry for small deformations.- }The new framework does capture the known (continuum) geometry when the latter applies. This is
illustrated in Fig.~1(c) for a profile $z(r)=A\exp[-r^2/\sigma^2]$ with $A$=0.8 \AA{} and $\sigma$=50 \AA{} \cite{deJuanPRB}. The continuum geometrical
invariants are: $\det(\tilde{g})=1+4r^2z^2/\sigma^4$, $\text{Tr}(\tilde{g})=2+4r^2z^2/\sigma^4$ ({both are radial-symmetric}),
$\tilde{K}=\frac{z'z''}{r(1+z'^2)^2}$, and $\tilde{H}=\frac{z'}{2r\sqrt{1+z'^2}}+\frac{z''}{2(1+z'^2)^{3/2}}$. We next lay out a crystalline graphene lattice
with discrete coordinates $(x_i,y_i,0)$, and assign $z_i\equiv z(r_i)$ to each atom, with $r_i=\sqrt{x_i^2+y_i^2}$. For easy comparison with the continuum
metric, we renormalize $g$ with respect to the flat discrete metric ($z_i=0$) [$\det(g_{(0)})=3a_0^4/4$, and Tr$(g_{(0)})/2=a_0^2$], and plot $1-\sqrt{\det{g}}$
and $2-\text{Tr}(g)$ to emphasize deviations from the reference metric. 
 $\Delta\sqrt{\det(g)}\equiv \sqrt{\det(\tilde{g})}-\sqrt{\det(g)}$ and $\Delta\text{Tr}(g)\equiv\text{Tr}(\tilde{g})-\text{Tr}(g)$ in Fig.~1(c) point to small discrepancies among the discrete ($g$) and continuum ($\tilde{g}$) \cite{Lee,doCarmo} metrics, originating already because $\tilde{g}$ is built from tangent vector fields $\tilde{\mathbf{g}}_i$ as two points along a continuum geodesic collapse onto each other, and this limiting process does not take place on the atomic lattice [see Fig.~\ref{fig:nodalpoints}(b)]. Those discrepancies aggravate
 under extreme morphologies for which the discrete geometry lacks smooth approximations in between atomic positions. On the other hand, $K_D=\tilde{K}$ and $H_D=\tilde{H}$ [Fig.~\ref{fig:nodalpoints}(c)],
 highlighting the meaning of curvature from atoms [Eqs.~(2-3)].

To make the discrete ($g$) and continuum ($\tilde{g}$) metrics correspond with one another, $\tilde{g}$ must be corrected at atomic positions as follows:
$g_{\alpha\beta}=b_{\alpha}^ib_{\beta}^j\tilde{g}_{ij}+b_{\alpha}^3b_{\beta}^3$,
where $\tilde{\mathbf{g}}_3=\hat{\tilde{\mathbf{n}}}=\tilde{\mathbf{g}}_1\times\tilde{\mathbf{g}}_2/|\tilde{\mathbf{g}}_1\times\tilde{\mathbf{g}}_2|$ and
$b_\alpha^k=\mathbf{a}_{\alpha}\cdot\tilde{\mathbf{g}}_k$ ($\alpha, \beta=1,2$, and $i,j,k=1,2,3$). The first term accounts for the
anisotropy of the honeycomb lattice, while the second one is an {\em exponential map} that brings continuum tangent fields
$\tilde{\mathbf{g}}_i$ back onto the atomistic surface \cite{Lee,M2,M4}.

\noindent{\em Rippled graphene.- }
The importance of a sound geometrical framework is motivated by rippled graphene. We contrast ripples created
by thermal fluctuations \cite{Fasolino1} with those created at low temperature due to edges. These two mechanisms lead to different types
of geometries (hence different magnitudes of strain-derived gauges). In a system with periodic boundary conditions, thermal fluctuations
create significant changes in interatomic distances (i.e., in the metric) \cite{Fasolino1}  and --as the boundaries are fixed-- such increases on interatomic
distances reflect on out-of-plane deformations (i.e., rippling).

Now consider a square graphene sample with about three million atoms, and relieve strain at the edges at the low temperature of 1 Kelvin.
The resulting membrane is shown in Fig.~2(a), where colors indicate varying heights across the sample \cite{usSSC}. Ripples in Ref.~\cite{Fasolino1} originate
from {\em increases} in the metric. On the other hand, the white margin in
between the ``rippled'' (curved) sample and the (yellow) exterior frame highlights an apparent {\em contraction} of our finite sample
when seen from above.

\begin{figure}[tb]
\begin{center}
\includegraphics[width=0.47\textwidth]{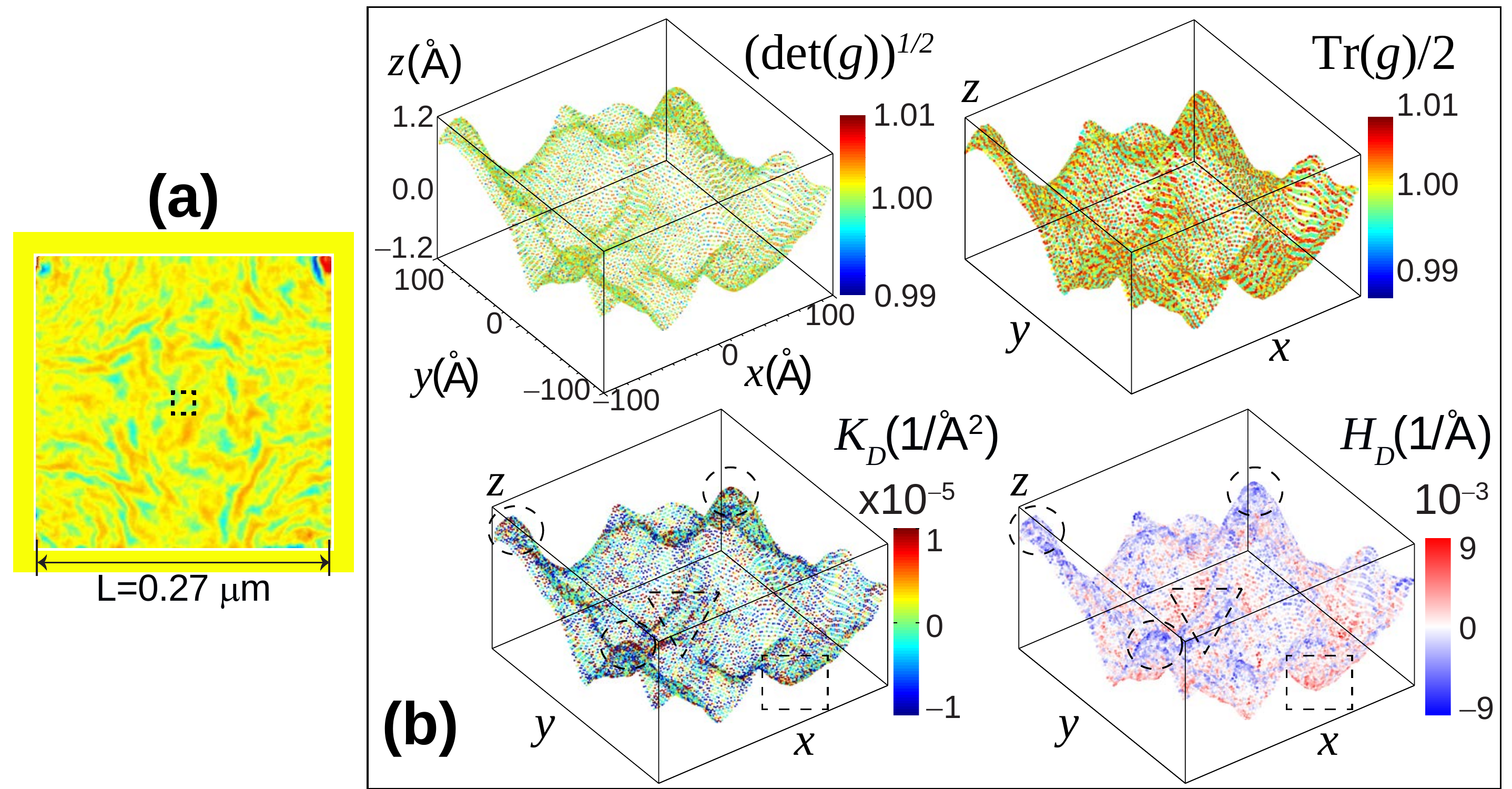}
\end{center}
\caption{(Color online) (a) Creation of ripples by cutting a square with side $L=0.27 \mu m$ at 1 Kelvin: The membrane
trades a planar configuration for a rippled one. (b) Geometrical invariants within the dashed square shown in (a). 
}\label{fig:f2}
\end{figure}

The details of this geometry are shown in Fig. 2(b). We find that det($g$) and Tr$(g)$ are unity
almost everywhere (yet there are significant random fluctuations driving the scales): The metric tells us that the membrane
does not contract and its area thus remains
almost unchanged. We show in Fig.~2(b) the discrete
curvatures, highlighting cusps by ovals, valleys by squares, and ridges by triangles.
Cusps and valleys have the largest Gauss curvature $K_D$ (deep red), while
ridges have the smallest one (deep blue). As expected, the mean curvature $H_D$ takes its largest (smallest) value at valleys (cusps) and alternates sign around
ridges. The curvature --without metric increases-- explains the white margins on Fig.~2(a).

 {\em The discrete geometry reflects the mechanism leading to ripple formation}: This highlights the virtues
 of a geometry originating from atoms. In particular, an accurate determination of $H_D$ is important since $H_D$ leads to spin diffusion in rippled
 graphene \cite{Ando2000,Huertas-Hernando,Katsnelson}. Though much has been said about ripples, no geometrical study with the accuracy and
 insight provided here exists. The geometrical invariants in Fig.~\ref{fig:f2}(b) are larger in magnitude than those in
Fig.~\ref{fig:nodalpoints}(c) --obtained by a smooth deformation from the reference ($z=0$) initial configuration.

 The starting point in the continuum theory is a flat metric $\delta_{\alpha\beta}$. There, a non-zero curvature directly leads to
 increases in interatomic distances [Eq.~(1)], and a non-zero height is directly identified with a non-zero strain-derived gauge. A question
 then arises whether the sample under study actually obeys Eq.~(1). The situation shown in Fig.~2 is a counterexample to the geometry inferred
 from Eq.~(1) because the metric is almost constant even though the height profile $z$ is clearly non-flat
 (for a pseudo-length-preserving distortion). Gauge fields
 for similar samples were reported in Ref.~\cite{usSSC}. Rippled graphene is an excellent example that shows how crucial it is to know the exact
 geometry on a case-by-case basis, and Fig.~2 represents the accurate geometrical characterization of rippled graphene down to the atomic level.

\begin{figure}[tb]
\begin{center}
\includegraphics[width=0.47\textwidth]{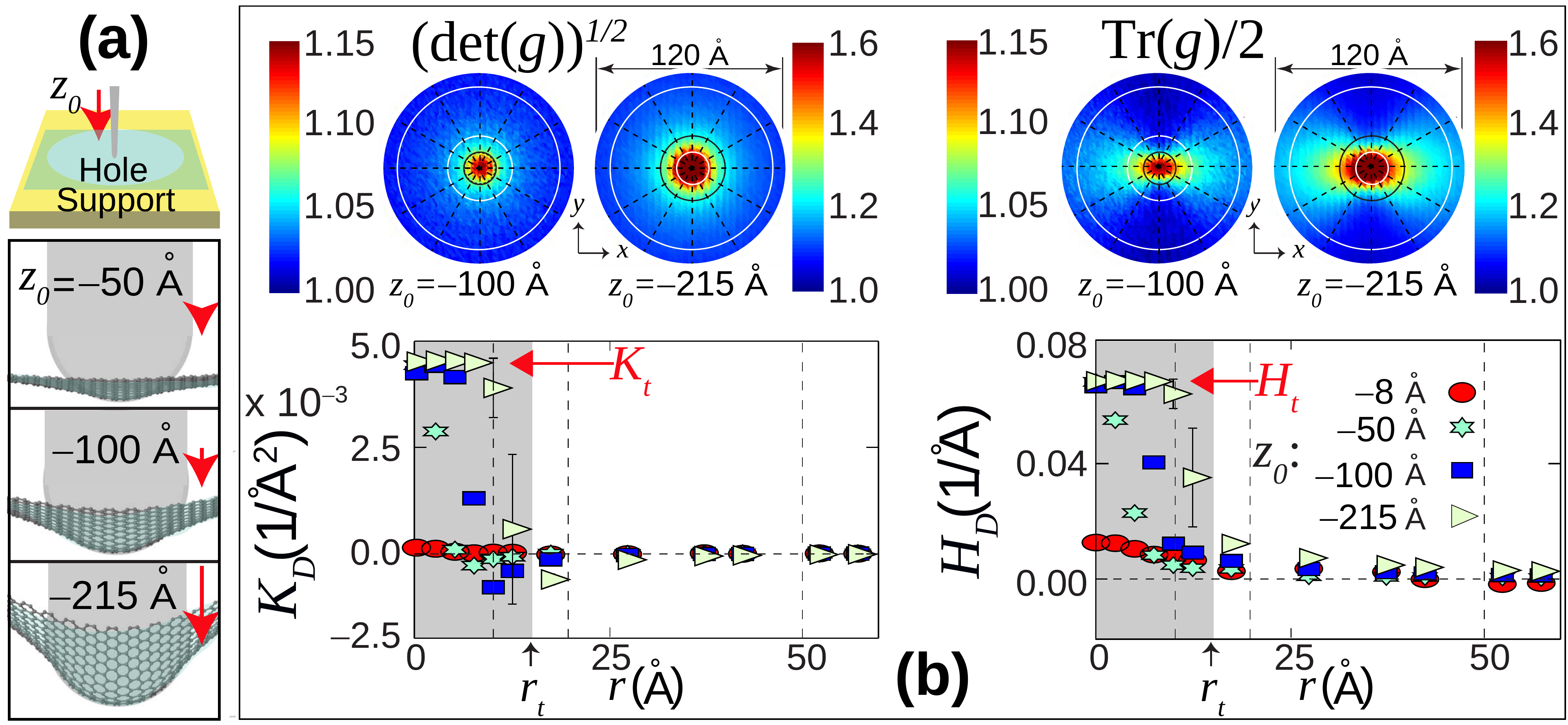}
\end{center}
\caption{(Color online) (a) Graphene under load by a semi-spherical tip of radius $r_t= 15$ \AA{}. (b) Geometrical invariants as the indentation proceeds.
Under load, the metric increases unbounded, yet {\em curvatures can only saturate}
to $K_t$ and $H_t$ as graphene conforms to the tip (see flat horizontal lines $K_D=K_t$ and $H_D=H_t$, for $0\le r\lesssim r_t$ at
$z_0=-100$ and $-215$ \AA{}).  Compare the trends with those in Fig.~1(c).}\label{fig:f3}
\end{figure}

\noindent{\em Graphene under load.- } Next we analyze a circular freestanding membrane \cite{stroscio} created by clamping the graphene sample in Fig.~2(a)
outside a radius $R=700$ \AA{} from the geometrical center. We push the membrane {\em down} to a depth $z_0$ with a spherical tip of radius
$r_t=15$ \AA{} [Fig.~3(a)]. The tip has constant curvatures $K_t\equiv 1/r_t^2=4.4\times 10^{-3}$ \AA$^{-2}$ and $H_t\equiv 1/r_t=0.07$ \AA$^{-1}$.

Figure 3(b) tells us quantitatively how graphene gradually conforms to the tip pushing it down. $g$ increases without bound (four upper plots in Fig.~3(b))
until an eventual mechanical breakdown \cite{Hone1}. $(\text{Tr}(g)/2)^{1/2}$ [from Fig.~\ref{fig:f3}(b)] informs of large increases of interatomic distances,
up to $\sim 26$\% for loads where $z_0=-215$ \AA{} \cite{Hone1,us}, beyond the realm of first-order continuum elasticity [Eq.~(1)]. The discrete metric $g$ also
uncovers an asymmetric elongation between armchair (vertical) and zigzag (horizontal) directions which the continuum metric $\tilde{g}$ does not capture unless corrected as indicated
above.

 On the other hand, graphene cannot acquire a curvature higher than that of the tip, so $K_D$ and $H_D$ must be bounded. This is precisely the
 content of the two lower plots in Fig.~3(b): For small loads ($z_0=-8$ \AA; ellipses) the curvatures are almost zero as expected. Curvatures
 increase (star, rectangle and triangle) as the magnitude of $z_0$ increases ($z_0=-50$, $-100$, and $-215$ \AA, respectively). The important observation
 is that curvatures saturate ($K_D\to K_t$ and $H_D\to H_t$) for distances $r$ within $r_t$ (shaded area), confirming the qualitative conformal shape
 depicted on Fig.~3(a). $K_D$ and $H_D$ have analogous trends in Figs.~1 and 3 [$H_D$ is a signed quantity, having opposite signs for a
 bulge (Fig.~1) and a sag (Fig.~3)]. Yet, it remains a challenge to accurately describe the geometry shown in Fig.~3 within the continuum formalism using
 Eq.~(1). This is so, because for high load the metric and curvature appear decoupled, while in the continuum approach they are inexorably inter-related.

\begin{figure}[tb]
\begin{center}
\includegraphics[width=0.47\textwidth]{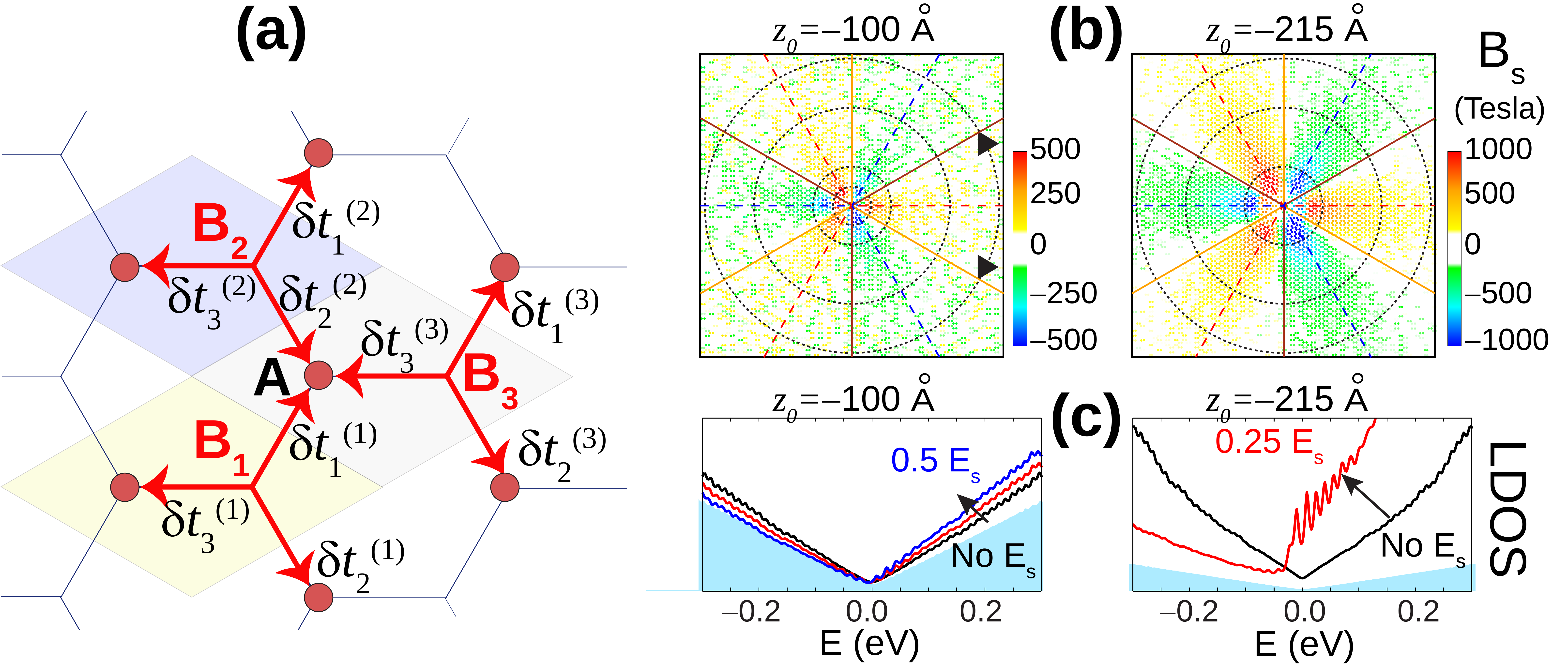}
\end{center}
\caption{(Color online) (a) The finite-difference curl leading to the pseudomagnetic field $B_s$ [Eq.~4] is obtained from hoppings among an
atom on the A-sublattice, and three neighboring atoms on B-sublattices. (b) $B_s$ for $z_0=-100$ and $-215$ \AA{} loads.
(c) LDOS with screened values of the deformation potential $E_s$ at $r=0$, for $z_0=-100$ and $-215$ \AA.{}}
\label{fig:f4}
\end{figure}

\noindent{\em Microscopic mass term and gauge fields.- }Next, we analyze electronic properties of graphene under load by the tip. We first
re-express the microscopic pseudo-magnetic field from the curl of the (pseudomagnetic) vector
potential. This is accomplished with a second-order difference relation among potential energies for an atom on the A-sublattice at the $K-$point [Fig.~4(a)]  \cite{RMP,Katsnelson,Manes,SI}:
\begin{eqnarray}\label{eq:mass}
-\mu_B B_s=&\frac{\sqrt{3}\hbar^2}{m_ea_0^2t}((\delta t_3^{(3)}-\delta t_1^{(3)})-
(\delta t_3^{(2)}-\delta t_1^{(2)})\\+&(\delta t_3^{(3)}-\delta t_2^{(3)})-(\delta t_3^{(1)}-\delta t_2^{(1)})).\nonumber
\end{eqnarray}
Here, $\mu_B$ is the Bohr magneton ($\simeq 5.8 \times 10^{-5}$ eV/Tesla), $\delta t_j^{(n)}$ is the standard change in hopping upon strain
at unit cell $n=1,2,3$ \cite{Ando2002,GuineaNatPhys2010,Vozmediano,us}, and $\frac{\sqrt{3}\hbar^2}{m_ea_0^2t}\simeq 2.5$ \cite{SI}. The pseudomagnetic field
$B_s$ changes sign at the B-sublattice and/or at the $K'$ point \cite{RMP,Katsnelson}. $E_s$ is the average deformation potential at a
given unit cell (see Refs.~\cite{us} and \cite{SI}) arising
 from the rearrangement of the electron cloud upon strain \cite{Ando2002}. $B_s$ is shown on Fig.~\ref{fig:f4}(b) for $z_0=-100$ and $z_0=-215$ \AA{} within a
 75 \AA{} radius from the tip.

\begin{figure}[tb]
\begin{center}
\includegraphics[width=0.47\textwidth]{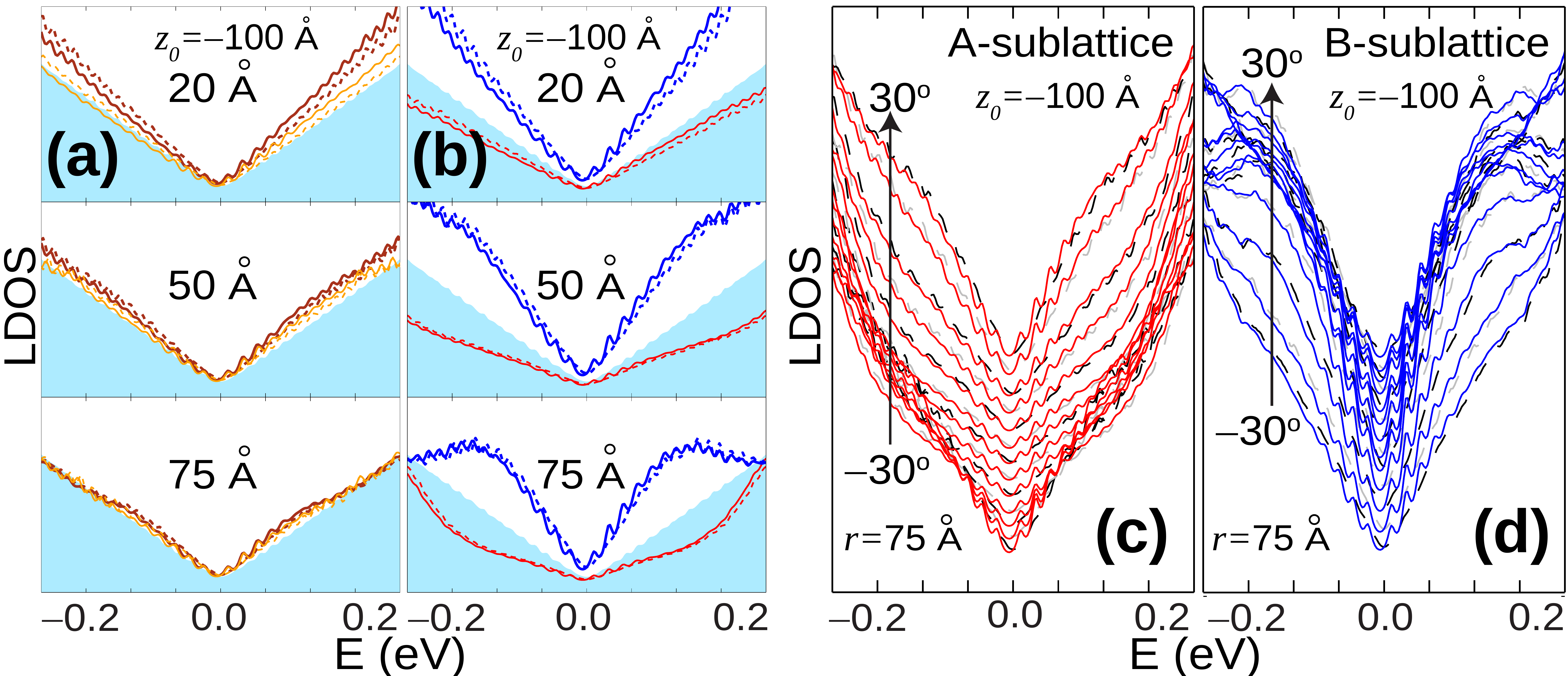}
\end{center}
\caption{(Color online) A-sublattice LDOS along the (a) $-30^o$ and (b) $+30^o$ axes ($z_0=-100$ \AA{}). Angular sweeps at $r=75$ \AA{} on the (c) A- and (d) B-sublattices ($z_0=-100$ \AA): Due to time-reversal symmetry, a sign change in $B_s$ in (b) has the same effect as a sublattice exchange (c-d).}
\label{fig:f5}
\end{figure}

\noindent{\em Local density of states.- }The shaded area plots on Fig.~\ref{fig:f4}(c) and Fig.~\ref{fig:f5}(a-b) are reference LDOS obtained from a
flat configuration with no strain. The metric and curvature in Fig.~3(b) take extreme values at $r=0$, and the LDOS in Fig.~4(c) increase in slope
as the Fermi velocity $v_F$ becomes more and more renormalized \cite{deJuanPRB,deJuanPRL2012} as $LDOS\propto 1/v_F^2$ \cite{RMP}. Remarkably, in
the scenario given by Fig.~3(b), the metric increase enhances the velocity renormalization while, at the same time, the curvature remains the same.
 This is so because the observed Fermi velocity renormalization is related to $g$: Indeed, it is caused by changes in interatomic
 distances \cite{Ando2002,Vozmediano}.

When a screened $E_s$ is applied, $v_F$ renormalization becomes electron-hole asymmetric \cite{RMP,MolecularGraphene}, and a sequence of equally-spaced peaks
arise even without explicit inclusion of spin (we do not have a quartet-splitting mechanism \cite{stroscio}). Thus, our results suggest an alternative
explanation for the identically-spaced LDOS features observed in a similar experimental setup \cite{stroscio} (in particular, refer to Fig.~4(c) with
$z_0=-215$ \AA{}, where $E_s$ is larger). We emphasize that there is no central LDOS peak (`zeroth Landau level') for circular membranes under load.

 The dashed LDOS curves in Fig.~5 were plotted with $E_s=0$, while the asymmetric LDOS profiles --displaying equally-spaced peaks-- were obtained
 with a (screened) $0.25E_s$ deformation potential. On Figs.~\ref{fig:f5}(a-b) we explore the LDOS in space and {\em exclusively} on the A-sublattice,
 generated from a $z_0=-100$ \AA{} load. Figure~5(a) shows the LDOS along the $-30^{o}$ (orange/light)  and  $+30^{o}$ (brown/dark) radial axes on the polar
 grid in display on Figs.~\ref{fig:f3}(c) and \ref{fig:f4}(b). Due to threefold-symmetry, the LDOS is identical upon $120^{o}$ rotations. $B_s\sim 0$ in
 Fig.~5(a) as it alternates sign at those axes. Hence, the only observable effect is a LDOS renormalization due to the metric \cite{deJuanPRB}
 ($r=20$ \AA{} plot). The renormalization gradually decreases with increasing $r$ until the LDOS overlaps with the reference one
 (see $r=75$ \AA{} plot), consistent with a metric approaching the flat one [Fig.~3(b)]. On the other hand, under a non-zero $B_s$ the LDOS on the A-sublattice
 either becomes enhanced (blue curves; $B_s<0$; $60^{o}$ axis) or suppressed (red; $B_s>0$; $0^{o}$ axis) with respect to the reference LDOS [Fig.~5(b)].
 Figure 4(a) complements previous reports \cite{Blanter}.

 Due to time-reversal symmetry, the A- and B-sublattices are subjected to $B_s$ with opposite signs and the behavior on Fig.~\ref{fig:f5}(b) should be
 reproducible by exploring the LDOS under the same $B_s$, but now at the B-sublattice (exchanging $K$ to $K'$ amounts to a sublattice exchange
 \cite{MolecularGraphene}). This is verified on Fig.~\ref{fig:f5}(c-d) by the sublattice resolved angular sweep across a $B_s>0$ feature. A sublattice
 asymmetric LDOS [Fig.~\ref{fig:f5}(c-d)] is consistent with a sublattice-dependent potential, Eq.~4, \cite{usSSC,Manes} through Coulomb's law.

\noindent{\em Conclusion.- }We presented a discrete approach to study graphene's geometry and its electron properties without relying on continuum
approximations and beyond thin-plate continuum mechanics. We used the method to study the experimentally relevant situations of rippled graphene
and graphene under large mechanical load. Our theory fully respects the discrete geometry of arbitrarily-shaped graphene, thus opening a completely
unexplored and promising route for strain-engineering beyond the restrictions of small and slowly-varying deformations inherent to continuum theories.  We
thank M.A.H. Vozmediano, D. Kennefick, and M. Mehboudi. We carried calculations at TACC ({\em Stampede}; TG-PHY090002) and Arkansas. M.~V. acknowledges
the Serbian Ministry of Science, Project 171027.

\noindent{\bf Supplementary Information:}

\noindent{\em Definition of metrics.- }Given two in-plane vector fields $\mathbf{g}_1$ and $\mathbf{g}_2$, metrics  $g_{\alpha\beta}\equiv\mathbf{g}_{\alpha}\cdot \mathbf{g}_{\beta}$ are symmetric ($g_{\alpha\beta}=g_{\beta\alpha}$) and positive-definite ($g_{\alpha\alpha}>0$) ($\alpha,$ $\beta=1,2$).\\

\noindent{\em The continuum geometry.- }Differential geometry and first-order continuum mechanics couple as:
\begin{equation}
g_{\alpha\beta}=\delta_{\alpha\beta}+2u_{\alpha\beta}; \text{ }k_{\alpha\beta}=\hat{\mathbf{n}}\cdot\mathbf{g}_{\alpha;\beta}\equiv\hat{\mathbf{n}}\cdot \frac{\partial \mathbf{g}_{\alpha}}{\partial \xi^{\beta}} +\Gamma^3_{\alpha\beta}.
\end{equation}
As it turns out, the connection $\Gamma^3_{\alpha\beta}$ is identically zero, leading to Equation 1 on the main manuscript.\\

\noindent{\em The realm of discrete differential geometry.- }The aims and scope of discrete differential geometry (DDG) are given here by adapting work of Bobenko and Suris \cite{Math2} to graphene's context.

The goal of DDG is to develop mathematically sound relations between differential and discrete geometry \cite{Math1,Math2}. Classical, {\em Riemannian} differential geometry studies the properties of smooth, continuum objects, and {\em discrete} geometry studies geometrical shapes made of polyhedra. DDG, in turn, seeks for discrete equivalents of notions and methods of continuous Riemannian geometry. Given that graphene's lattice is made of polyhedra, it represents a physically-relevant system for DDG.

 To realize theories consistent with DDG one first determines a proper discrete surface, and develops theory {\em from that discrete surface}. In the absence of an actual atomic lattice, one can suggest many different discretizations of surfaces having the same continuum limit. For graphene, on the other hand, the honeycomb lattice {\em is} the discrete lattice, and no more fundamental choice exists without involving approximations.

This represents a central difference between an all-discrete theory and  discrete approximations of continuum models. In the latter, discretization of surfaces and differential equations is carried out on an arbitrary mesh. In the context of strain engineering in graphene, this process starts the moment the theory of an effective continuum media \cite{Ando2002,GuineaNatPhys2010,Vozmediano} is mapped onto arbitrary meshes for numerical analysis. In applying DDG to graphene, on the other hand, the mesh is given by the deformed honeycomb lattice, and we never take the continuum limit of the pseudospin Hamiltonian when considering the electronic behavior either \cite{us,usSSC}. The results of DDG are therefore non-perturbative on graphene's atomistic morphology.\\

\noindent{}{\em Discrete geometry for substantial distortions.- }On page 2 of the main manuscript we demonstrate that the discrete and continuum geometries agree in the limit of small deformations; this is, when the distortion is small when compared to interatomic distances. For large deformations, the continuum hypothesis breaks down as continuum tangent fields cannot be generated to arbitrary precision from atomic locations.

 To show the breakdown of the continuum geometrical description, we display in Fig.~1 the geometry for the function $z(x_i,y_i)=A(\exp[-\frac{(x_i-x_0)^2+y_i^2}{\sigma^2}]-\exp[-\frac{(x_i+x_0)^2+y_i^2}{\sigma^2}])$, with $A=25$ \AA, $\sigma=30$ \AA, and $x_0=25$ \AA{}. Here, the continuum geometrical invariants display large discrepancies to the discrete geometry; this is particularly the case for the trace of the metric.

\begin{figure*}[tb]
\begin{center}
\includegraphics[width=0.9\textwidth]{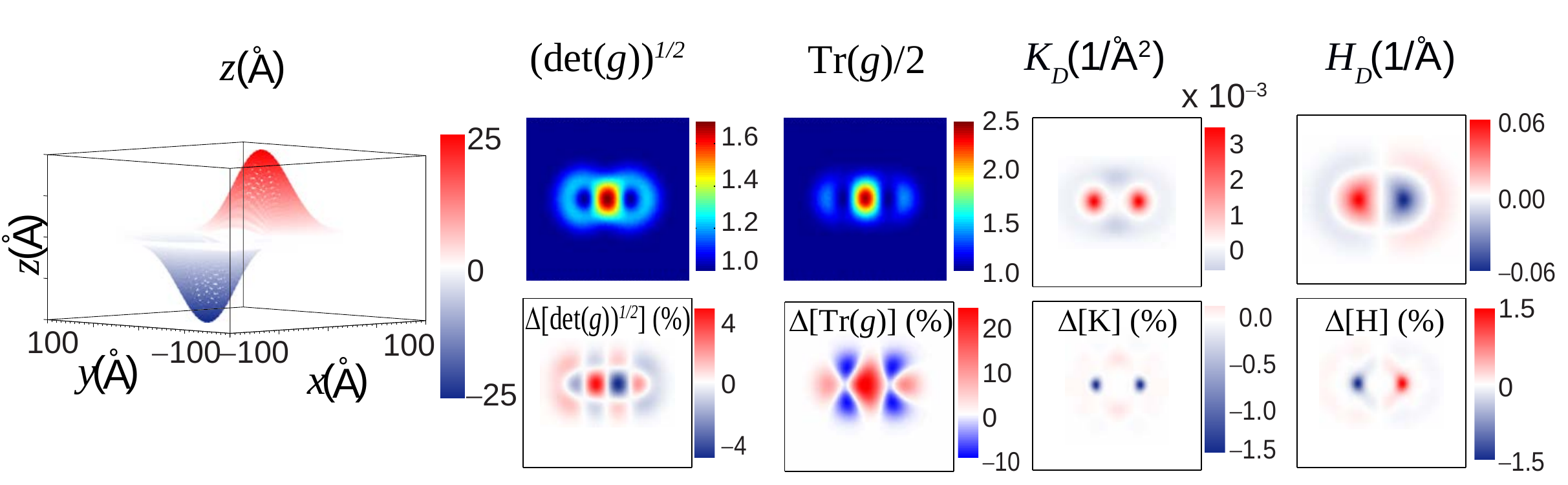}
\end{center}
\caption{(Color online) Discrepancies among the continuum and discrete geometries for a large deformation.}\label{fig:SI-m1}
\end{figure*}

 As a consequence of such discrepancies, the continuum idealization of the atomic membrane will have an inaccurate distribution of in-plane local forces/stress, and will be incompatible with the actual lattice structure. A subtler deficiency, tangent vectors $\tilde{\mathbf{g}}_i$ ($i=1,2$) may {\em lie outside} of the polygonal surface [Fig.~1(b) on the main text], compromising mechanical equilibrium. An exponential map bringing the continuum vector field $\tilde{\mathbf{g}}_i$ back onto the atomistic surface becomes necessary, and the continuum metric $\tilde{g}$ must be corrected at atomic positions to properly conform to $g$:
\begin{equation}\label{eq:correction}
g_{\alpha\beta}=b_{\alpha}^ib_{\beta}^j\tilde{g}_{ij}+b_{\alpha}^3b_{\beta}^3,
\end{equation}
as indicated on Page 2 of the main manuscript. The first term on Eq.~\eqref{eq:correction} accounts for the anisotropy of the atomic distortion and the second one is the exponential map. In looking at Figure 1, it is important to note that curvatures are much more better behaved in between descriptions, highlighting once again the deep significance of curvature from atoms, Eqns.~2 and 3 on the main text.\\

\begin{figure}[h]
\begin{center}
\includegraphics[width=0.48\textwidth]{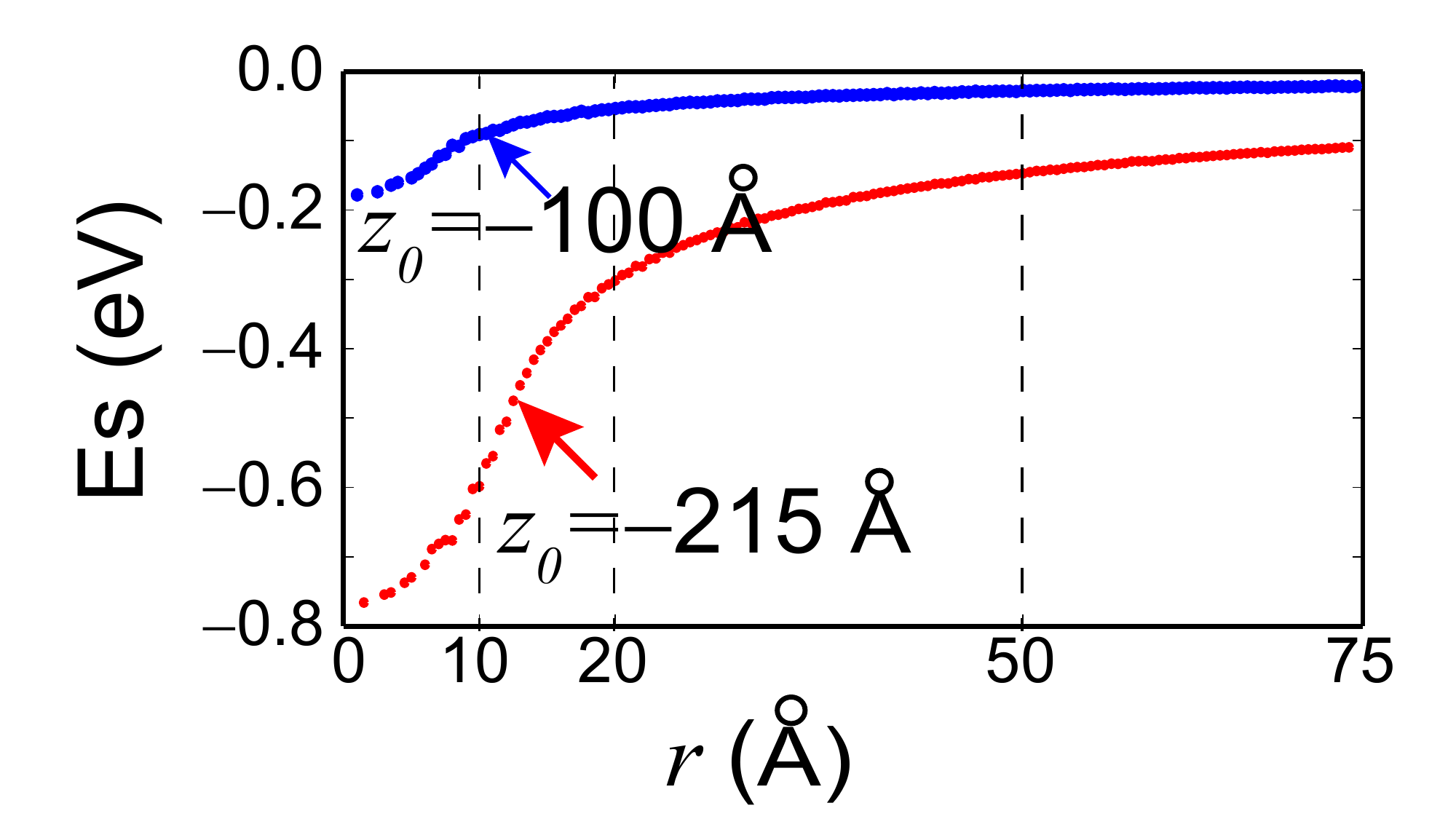}
\end{center}
\caption{(Color online) Radial profile of the deformation potential $E_s$.}\label{fig:SI0}
\end{figure}

\noindent{}{\em The mean deformation potential $E_s$.- }This empirical expression was introduced before (Eqn.~(18) in Ref.~\cite{us}), where we proposed it to be linearly proportional to the average increases in bond lengths. $E_s$ is radially-symmetric, and it has the radial shape illustrated in Fig.~\ref{fig:SI0} for loads $z_0=-100$ \AA, and $z_0=-215$ \AA.

\noindent{}{\em Derivation of the pseudo-magnetic field.- }Changes in distances between atoms upon strain modify the local electrostatic potential \cite{usSSC}. Proper consideration of those changes on the local potential cannot be given by consideration of nearest neighbors alone, and a second-order difference equation is needed. Here we derive microscopic expressions for the pseudo-magnetic field, when the zigzag direction is parallel to the y-axis \cite{us}. Results for the other common choice (i.e., zigzag direction parallel to x-axis \cite{GuineaNatPhys2010,Vozmediano}) can be  obtained along similar lines.

\begin{figure}[h]
\begin{center}
\includegraphics[width=0.48\textwidth]{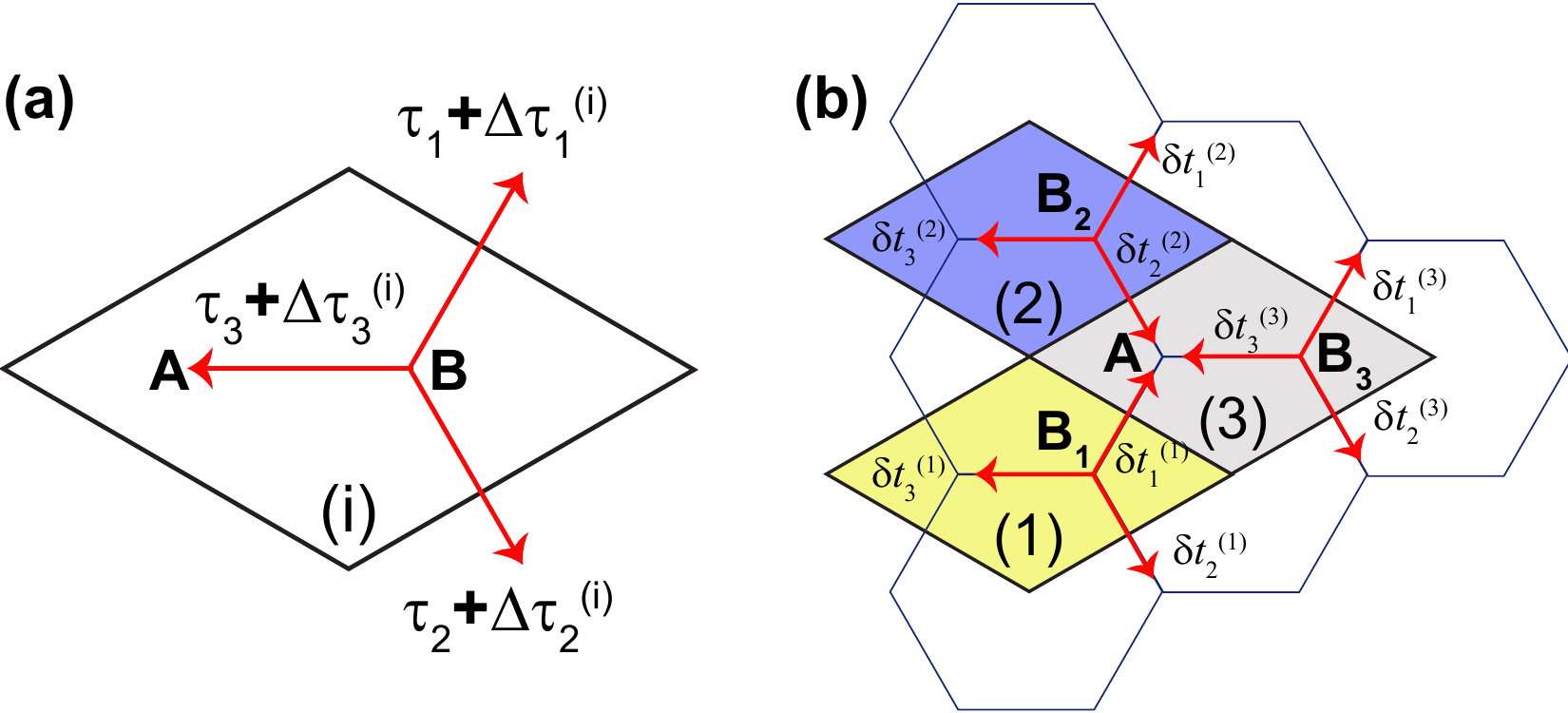}
\end{center}
\caption{(Color online) Schematics of local deformations required in determining the magnitude of the pseudo-magnetic field by finite differences.}\label{fig:SI1}
\end{figure}

The program of action is as follows: We wish to express the magnitude of the finite-difference pseudo-magnetic field $B_s$ as a function of local changes on $\delta t$ in between sublattices from terms leading to the vector potential. $B_s$ will display a straightforward and physically intuitive form in terms of energy variations $\delta t$ among neighboring atoms belonging to complementary sublattices. These variations are similar in origin to the ones we reported before \cite{usSSC}, but with the symmetry of a Zeeman term \cite{Manes} built in.

 We set the zigzag direction to be parallel to the y-axis, and start with the standard expression leading to the vector potential:
\begin{equation}
\sum_{j=1}^3\delta t_je^{i\mathbf{K}\cdot \boldsymbol{\tau}_j},
\end{equation}
where $\boldsymbol{\tau}_1=(1/2,\sqrt{3}/2)a_0/\sqrt{3}$, $\boldsymbol{\tau}_2=(1/2,-\sqrt{3}/2)a_0/\sqrt{3}$, and $\boldsymbol{\tau}_3=(-1,0)a_0/\sqrt{3}$.
$\delta t_j=-|\beta|t\boldsymbol{\tau}_j\cdot\Delta\boldsymbol{\tau}_j/a_0^2$, see \cite{Vozmediano,us} for details. We choose $\mathbf{K}=(0,1)\frac{4\pi}{3a_0}$ as well. Then:
\begin{eqnarray}\label{eq:functions}
\sum_{j=1}^3\delta t_j^{(n)}e^{i\mathbf{K}\cdot \boldsymbol{\tau}_j}=\\
-\frac{\sqrt{3}}{2}\left[
\frac{2\delta t_3^{(n)}-\delta t_1^{(n)}-\delta t_2^{(n)}}{\sqrt{3}}+i(\delta t_1^{(n)}-\delta t_2^{(n)})\nonumber
\right].
\end{eqnarray}
The upper index $(n)$ enters in Eq.~\eqref{eq:functions} because the discrete curl (a term arising from differences of $\delta t$ in between sublattices) requires obtaining differences of $\sum_{j=1}^3\delta t_j^{(n)}e^{i\mathbf{K}\cdot \boldsymbol{\tau}_j}$ on three adjacent unit cells Fig.~\ref{fig:SI1}; $n=1,2,3$. Consistent with the choice of zigzag direction, the $x$ and $y$ components of the vector potential dictate the choice of components in Eq.~\eqref{eq:functions} \cite{us}. To simplify the algebra, we introduce:
\begin{eqnarray}\label{eq:fs}
f_x(A)\equiv   &\delta t_2^{(2)}-\delta t_1^{(1)}, \text{ } f_y(A)\equiv\frac{2\delta t_3^{(3)}-\delta t_1^{(1)}-\delta t_2^{(2)}}{\sqrt{3}},\nonumber\\
f_x(B_1)\equiv &\delta t_2^{(1)}-\delta t_1^{(1)}, \text{ } f_y(B_1)\equiv\frac{2\delta t_3^{(1)}-\delta t_1^{(1)}-\delta t_2^{(1)}}{\sqrt{3}},\nonumber\\
f_x(B_2)\equiv &\delta t_2^{(2)}-\delta t_1^{(2)}, \text{ } f_y(B_2)\equiv\frac{2\delta t_3^{(2)}-\delta t_1^{(2)}-\delta t_2^{(2)}}{\sqrt{3}},\nonumber\\
f_x(B_3)\equiv &\delta t_2^{(3)}-\delta t_1^{(3)}, \text{ } f_y(B_3)\equiv\frac{2\delta t_3^{(3)}-\delta t_1^{(3)}-\delta t_2^{(3)}}{\sqrt{3}}.
\end{eqnarray}
We use Eq.~\eqref{eq:fs} to establish the discrete local curl in terms of differences of function $\mathbf{f}=(f_x,f_y)$ at neighboring positions corresponding to complementary sublattices:
\begin{equation}\label{eq:curl}
\Delta\times \mathbf{f}\equiv(\Delta_x f_y-\Delta_y f_x)\hat{\mathbf{n}}.
\end{equation}
Then, finite-differences become:
\begin{eqnarray}\label{eq:dxfy}
\Delta_x f_y&=\frac{f_y(A)-f_y(B_3)}{(\boldsymbol{\tau}_3+\Delta\boldsymbol{\tau}_3^{(3)})\cdot \hat{\mathbf{i}}}+
\frac{f_y(A)-f_y(B_2)}{(\boldsymbol{\tau}_2+\Delta\boldsymbol{\tau}_2^{(2)})\cdot \hat{\mathbf{i}}}+
\frac{f_y(A)-f_y(B_1)}{(\boldsymbol{\tau}_1+\Delta\boldsymbol{\tau}_1^{(1)})\cdot \hat{\mathbf{i}}},
\end{eqnarray}
and:
\begin{eqnarray}\label{eq:dyfx}
\Delta_y f_x&=\frac{f_x(A)-f_x(B_2)}{(\boldsymbol{\tau}_2+\Delta\boldsymbol{\tau}_2^{(2)})\cdot \hat{\mathbf{j}}}+
\frac{f_x(A)-f_x(B_1)}{(\boldsymbol{\tau}_1+\Delta\boldsymbol{\tau}_1^{(1)})\cdot \hat{\mathbf{j}}}.
\end{eqnarray}
$\hat{\mathbf{i}}$ and $\hat{\mathbf{j}}$ represent local in-plane vector fields:
\begin{equation}
\hat{\mathbf{i}}\equiv \frac{\mathbf{a}_1+\mathbf{a}_2}{|\mathbf{a}_1+\mathbf{a}_2|}; \text{ }
\hat{\mathbf{j}}\equiv \frac{\mathbf{a}_1-\mathbf{a}_2}{|\mathbf{a}_1-\mathbf{a}_2|},
\end{equation}
with
$\mathbf{a}_1=\boldsymbol{\tau}_1+\Delta\boldsymbol{\tau}_1^{(3)}-(\boldsymbol{\tau}_3+\Delta\boldsymbol{\tau}_3^{(3)})$ and
$\mathbf{a}_2=\boldsymbol{\tau}_2+\Delta\boldsymbol{\tau}_2^{(3)}-(\boldsymbol{\tau}_3+\Delta\boldsymbol{\tau}_3^{(3)})$
 the local lattice displacements for a central atom on the $A-$sublattice, and $\hat{\mathbf{n}}$ was defined in Eq.~2 in the main text. Using Eq.~\eqref{eq:fs}, we get for Eqs.~\eqref{eq:dxfy} and \eqref{eq:dyfx}:
\begin{eqnarray}
\Delta_xf_y=\frac{(\delta t_1^{(3)}+\delta t_2^{(3)})-(\delta t_1^{(1)}+\delta t_2^{(2)})}{(\boldsymbol{\tau}_3+\Delta\boldsymbol{\tau}_3^{(3)})\cdot \hat{\mathbf{i}}}+\nonumber\\
\frac{2\delta t_3^{(3)}-\delta t_1^{(1)}-(2\delta t_3^{(2)}-\delta_1^{(2)})}{(\boldsymbol{\tau}_2+\Delta\boldsymbol{\tau}_2^{(2)})\cdot \hat{\mathbf{i}}}+\nonumber\\
\frac{2\delta t_3^{(3)}-\delta t_2^{(2)}-(2\delta t_3^{(1)}-\delta_2^{(1)})}{(\boldsymbol{\tau}_1+\Delta\boldsymbol{\tau}_1^{(1)})\cdot \hat{\mathbf{i}}},
\end{eqnarray}
and:
\begin{eqnarray}\label{eq:dyfx_}
\Delta_y f_x=\frac{\delta t_1^{(2)}-\delta t_1^{(1)}}{(\boldsymbol{\tau}_2+\Delta\boldsymbol{\tau}_2^{(2)})\cdot \hat{\mathbf{j}}}+
\frac{\delta t_2^{(2)}-\delta t_2^{(1)}}{(\boldsymbol{\tau}_1+\Delta\boldsymbol{\tau}_1^{(1)})\cdot \hat{\mathbf{j}}}.
\end{eqnarray}

Next, we require hermiticity of a vector potential (these conditions are not needed for a scalar potential term, which is Hermitian by construction). The conditions are (see \cite{GuineaNatPhys2010} and \cite{us,usSSC} for extended discussion):
\begin{equation}\label{eq:conditions}
\boldsymbol{\tau}_1+\Delta\boldsymbol{\tau}_1^{(1)}\to \boldsymbol{\tau}_1+\Delta\boldsymbol{\tau}_1^{(3)}, \text{ and }
\boldsymbol{\tau}_2+\Delta\boldsymbol{\tau}_2^{(2)}\to \boldsymbol{\tau}_2+\Delta\boldsymbol{\tau}_2^{(3)}.
\end{equation}
An immediate consequence from Eq.~\eqref{eq:conditions} is that $\delta t_1^{(1)}\to \delta t_1^{(3)}$ and $\delta t_2^{(2)}\to \delta t_2^{(3)}$ as well. Therefore, Eqs.~\eqref{eq:dxfy} and \eqref{eq:dyfx} take the final form:
\begin{eqnarray}\label{eq:dxfy2}
\Delta_xf_y=\frac{2\delta t_3^{(3)}-\delta t_1^{(4)}-(2\delta t_3^{(2)}-\delta_1^{(2)})}{(\boldsymbol{\tau}_2+\Delta\boldsymbol{\tau}_2^{(3)})\cdot \hat{\mathbf{i}}}+\nonumber\\
\frac{2\delta t_3^{(3)}-\delta t_2^{(3)}-(2\delta t_3^{(1)}-\delta_2^{(1)})}{(\boldsymbol{\tau}_1+\Delta\boldsymbol{\tau}_1^{(3)})\cdot \hat{\mathbf{i}}},
\end{eqnarray}
and
\begin{eqnarray}\label{eq:dyfx2}
\Delta_y f_x=\frac{\delta t_1^{(2)}-\delta t_1^{(3)}}{(\boldsymbol{\tau}_2+\Delta\boldsymbol{\tau}_2^{(3)})\cdot \hat{\mathbf{j}}}+
\frac{\delta t_2^{(3)}-\delta t_2^{(1)}}{(\boldsymbol{\tau}_1+\Delta\boldsymbol{\tau}_1^{(3)})\cdot \hat{\mathbf{j}}}.
\end{eqnarray}
Equation~\eqref{eq:curl} as derived here supersedes our previous expression for the curl leading to the pseudo-magnetic field \cite{us}. Equation~\eqref{eq:curl} together with Eqs.~\eqref{eq:dxfy2} and \eqref{eq:dyfx2} were employed in plotting Fig.~4(b) in the main text.

 The following approximation helps the reader in better grasping the origin of the curl from differences of changes of on-site potentials upon strain (i.e., a `second-order' differences equation). If we set $\frac{1}{(\boldsymbol{\tau}_j+\Delta\boldsymbol{\tau}_j^{(n)})\cdot\hat{\mathbf{i}}}\simeq
 \frac{1}{\boldsymbol{\tau}_j\cdot\hat{\mathbf{i}}}$ (and a similar approximation for the term involving projection onto $\hat{\mathbf{j}}$), then the vector projections on the denominator can be carried out easily, and the finite-differences curl takes the following rather intuitive form:
\begin{eqnarray}
|\Delta \times \mathbf{f}|=\Delta_xf_y-\Delta_yf_x=\frac{4}{a_0}[(\delta t_3^{(3)}-\delta t_1^{(3)})\\
+(\delta t_3^{(3)}-\delta t_2^{(3)})+(\delta t_1^{(2)}-\delta t_3^{(2)})+(\delta t_2^{(1)}-\delta t_3^{(1)})].\nonumber
\end{eqnarray}
This way:
\begin{eqnarray}
\mathbf{B}_s=-2\sqrt{3}\frac{\phi_0}{\pi a_0^2 t}\times \nonumber\\
((\delta t_3^{(3)}-\delta t_1^{(3)})+(\delta t_3^{(3)}-\delta t_2^{(3)})+\nonumber\\
(\delta t_1^{(2)}-\delta t_3^{(2)})+(\delta t_2^{(1)}-\delta t_3^{(1)}))\hat{\mathbf{n}},
\end{eqnarray}
($\hat{\mathbf{n}}=\hat{z}$ is the local normal; all constants were defined before \cite{us}; $a_0=1.391$ at 1 Kelvin), and the curl finally becomes Eq.~4 in the main text:
\begin{eqnarray}
-\mu_B B_s=2\sqrt{3}g \frac{e\hbar}{4m_e}\frac{\phi_0}{\pi a_0^2 t}\times((\delta t_3^{(3)}-\delta t_1^{(3)})+\nonumber\\
(\delta t_3^{(3)}-\delta t_2^{(3)})+(\delta t_1^{(2)}-\delta t_3^{(2)})+(\delta t_2^{(1)}-\delta t_3^{(1)})).
\end{eqnarray}
$g$ is the Land{\'e} factor, which we set equal to 2. The Land{\'e} factor $g$, along with $m_e$ --the electron mass-- are external parameters of the theory. The prefactor is dimensionless:
\begin{equation}
2\sqrt{3}g \frac{e\hbar}{4m_e}\frac{\phi_0}{\pi a_0^2 t}=2\sqrt{3}\frac{g\hbar^2}{4m_ea_0^2t}\simeq 2.5.
\end{equation}
$B_s$ changes sign upon sublattice exchange or $K-$point exchange. Eq.~(15) is related to the staggered DOS observed experimentally, giving additional insight as to how the discrete geometry of graphene couples to its electronic properties.\\

\noindent{}{\em Consequences for spin-orbit coupling estimates.- }The concepts presented in the main text can be used for an accurate determination of the spin-orbit coupling $\Delta(\mathbf{r}_i)$ induced by curvature. In the absence of strain $\Delta(\mathbf{r}_i)\propto  H_D(\mathbf{r}_i)$  \cite{Ando2000,Huertas-Hernando}. An estimate from $H_D$ in Fig.~2(c) in the main text yields $-0.2$ meV$\lesssim\Delta\lesssim 0.2$ meV \cite{Huertas-Hernando}. We note that $\Delta(\mathbf{r}_i)$ {\em changes sign with $H_D$}. The proportionality between $H_D$ and $\Delta$ depends on hopping invariants that decay exponentially with distance: A more general expression for $\Delta(\mathbf{r}_i)$ must depend on the metric $g$ as well.

 $H_D$ increases by an order of magnitude and changes sign in between Figs.~2(c) and 3(c) in the main text, making $|\Delta|\lesssim 2$ meV under central load --even when ignoring effects due to $g$. As we employed a smearing parameter $\sigma=5$ meV in plotting LDOS curves, we were unable to resolve spin-obit coupling, which was hence ignored in the main text.




\begin{thebibliography}{65}
\expandafter\ifx\csname natexlab\endcsname\relax\def\natexlab#1{#1}\fi
\expandafter\ifx\csname bibnamefont\endcsname\relax
  \def\bibnamefont#1{#1}\fi
\expandafter\ifx\csname bibfnamefont\endcsname\relax
  \def\bibfnamefont#1{#1}\fi
\expandafter\ifx\csname citenamefont\endcsname\relax
  \def\citenamefont#1{#1}\fi
\expandafter\ifx\csname url\endcsname\relax
  \def\url#1{\texttt{#1}}\fi
\expandafter\ifx\csname urlprefix\endcsname\relax\def\urlprefix{URL }\fi
\providecommand{\bibinfo}[2]{#2}
\providecommand{\eprint}[2][]{\url{#2}}
\bibitem[{\citenamefont{Wallace}(1947)}]{Wallace}
\bibinfo{author}{\bibfnamefont{P.}~\bibnamefont{Wallace}},
  \bibinfo{journal}{Phys. Rev.} \textbf{\bibinfo{volume}{71}},
  \bibinfo{pages}{622} (\bibinfo{year}{1947}); \bibinfo{author}{\bibfnamefont{A.~K.}~\bibnamefont{Geim}} and K.~S. Novoselov,
  \bibinfo{journal}{Nature Materials} \textbf{\bibinfo{volume}{6}}, \bibinfo{pages}{183}
  (\bibinfo{year}{2007}); \bibinfo{author}{\bibfnamefont{Y.}~\bibnamefont{Zhang}}, Y.-W. Tan, H.~L. Stormer and P. Kim,
  \bibinfo{journal}{Nature} \textbf{\bibinfo{volume}{438}},
  \bibinfo{pages}{201} (\bibinfo{year}{2005}); \bibinfo{author}{\bibfnamefont{C.}~\bibnamefont{Berger}},
  \bibnamefont{Z. Song, T. Li, X. Li, A.~Y. Ogbazghi, R. Feng, Z. Dai, A.~N. Marchenkov, E.~H. Conrad, P.~N. First, and W.~A. de Heer},
  \bibinfo{journal}{J. Phys. Chem. B}
  \textbf{\bibinfo{volume}{108}}, \bibinfo{pages}{19912}
  (\bibinfo{year}{2004}).

\bibitem{RMP}
   \bibinfo{author}{\bibfnamefont{A.~H.} \bibnamefont{{Castro-Neto}}}, F. Guinea, N.~M.~R. Peres, K.~S. Novoselov and A.~K. Geim,
  \bibinfo{journal}{Rev. Mod. Phys.} \textbf{\bibinfo{volume}{81}},
  \bibinfo{pages}{109} (\bibinfo{year}{2009}).

\bibitem{Novo}
  \bibinfo{author}{\bibfnamefont{K.~S.} \bibnamefont{{Novoselov}}}, D. Jiang, F. Schedin, T.~J. Booth, V.~V. Khotkevich,
S.~V. Morozov and A.~K. Geim,
  \bibinfo{journal}{Proc. Natl. Acad. Sci. (USA)} \textbf{\bibinfo{volume}{102}},
  \bibinfo{pages}{10451} (\bibinfo{year}{2005});
     \bibinfo{author}{\bibfnamefont{A.~K.} \bibnamefont{{Geim}}} and I.~V. Grigorieva,
  \bibinfo{journal}{Nature} \textbf{\bibinfo{volume}{499}},
  \bibinfo{pages}{419} (\bibinfo{year}{2013});
     \bibinfo{author}{\bibfnamefont{S.~Z.} \bibnamefont{{Butler}}}, {S.~M.} Hollen, L. Cao, Y. Cui, {J.~A.} Gupta, {H.~R.} Guti{\'e}rrez,
   {T.~F.} Heinz, {S.~S.} Hong, J. Huang, {A.~F.} Ismach, E. Johnston-Halperin, M. Kuno, {V.~V.} Plashnitsa,
   {R.~D.} Robinson, {R.~S.} Ruoff, S. Salahuddin, J. Shan, L. Shi, {M.~G.} Spencer, M. Terrones, W. Windl and {J.~E.} Goldberger
  \bibinfo{journal}{ACS Nano} \textbf{\bibinfo{volume}{7}},
  \bibinfo{pages}{2898} (\bibinfo{year}{2013});
     \bibinfo{author}{\bibfnamefont{H.~L.} \bibnamefont{{Zhuang}}}, A.~K. Singh and R.~G. Hennig,
  \bibinfo{journal}{Phys. Rev. B} \textbf{\bibinfo{volume}{87}},
  \bibinfo{pages}{165415} (\bibinfo{year}{2013}).

\bibitem[{\citenamefont{Morozov et~al.}(2006)\citenamefont{Morozov, Novoselov,
  Katsnelson, Schedin, Ponomarenko, Jiang, and Geim}}]{PRLMorozov2006}
\bibinfo{author}{\bibfnamefont{S.}~\bibnamefont{Morozov}}, K.~S. Novoselov, M.~I. Katsnelson, F. Schedin, L.~A. Ponomarenko, D. Jiang and A.~K. Geim,
  \bibinfo{journal}{Phys. Rev. Lett.} \textbf{\bibinfo{volume}{97}},
  \bibinfo{pages}{016801} (\bibinfo{year}{2006}); \bibinfo{author}{\bibfnamefont{M.~I.}~\bibnamefont{Katsnelson}} and A.~K. Geim,
  \bibinfo{journal}{Phil. Trans. R. Soc. A} \textbf{\bibinfo{volume}{366}},
  \bibinfo{pages}{195} (\bibinfo{year}{2008});
  \bibinfo{author}{\bibfnamefont{N.}~\bibnamefont{Mohanty}}, M. Fahrenholtz, A. Nagaraja, D. Boyle and V. Berry,
  \bibinfo{journal}{Nano Lett.} \textbf{\bibinfo{volume}{11}},
  \bibinfo{pages}{1270} (\bibinfo{year}{2011});
  \bibinfo{author}{\bibfnamefont{M.}~\bibnamefont{Yamamoto}}, O. Pierre-Louis, J. Huang, M.~S. Fuhrer, T.~L. Einstein, and W.~G. Cullen,
  \bibinfo{journal}{Phys. Rev. X} \textbf{\bibinfo{volume}{2}},
  \bibinfo{pages}{041018} (\bibinfo{year}{2012}).

\bibitem[{\citenamefont{Meyer et~al.}(2007)\citenamefont{Meyer, Geim,
  Katsnelson, Novoselov, Booth, and Roth}}]{Nature2007}
\bibinfo{author}{\bibfnamefont{J.~C.} \bibnamefont{Meyer}}, A.~K. Geim, M.~I. Katsnelson, K.~S. Novoselov, T.~J. Booth and S. Roth,
  \bibinfo{journal}{Nature} \textbf{\bibinfo{volume}{446}}, \bibinfo{pages}{60}
  (\bibinfo{year}{2007});
\bibinfo{author}{\bibfnamefont{M.}~\bibnamefont{Gass}}, U. Bangert, A.~L. Bleloch, P. Wang, R.~R. Nair and A.~K. Geim,
\bibinfo{journal}{Nature Nanotechnology} \textbf{\bibinfo{volume}{3}},
  \bibinfo{pages}{676} (\bibinfo{year}{2008}).


\bibitem[{\citenamefont{Tapaszto et~al.}(2012)\citenamefont{Tapaszto,
  Dumitrica, Kim, Nemes-Incze, Hwang, and Biro}}]{Biro}
\bibinfo{author}{\bibfnamefont{L.}~\bibnamefont{Tapaszto}}, T. Dumitrica, S.~J. Kim, P. Nemes-Incze, C. Hwang and L.~P. Biro,
  \bibinfo{journal}{Nature Phys.} \textbf{\bibinfo{volume}{8}},
  \bibinfo{pages}{739} (\bibinfo{year}{2012}).

\bibitem[{\citenamefont{Shenoy et~al.}(2008)\citenamefont{Shenoy, Reddy,
  Ramasubramaniam, and Zhang}}]{Shenoy}
\bibinfo{author}{\bibfnamefont{V.}~\bibnamefont{Shenoy}}, C.~D. Reddy, A. Ramasubramaniam and Y.~W. Zhang,
  \bibinfo{journal}{Phys. Rev. Lett.} \textbf{\bibinfo{volume}{101}},
  \bibinfo{pages}{245501} (\bibinfo{year}{2008}).

\bibitem[{\citenamefont{Huang et~al.}(2009)\citenamefont{Huang, Liu, Su, Wu,
  Duan, Gu, and Liu}}]{Liu}
\bibinfo{author}{\bibfnamefont{B.}~\bibnamefont{Huang}}, M. Liu, N. Su, J. Wu, W. Duan, B.-L. Gu and F. Liu,
  \bibinfo{journal}{Phys. Rev. Lett.} \textbf{\bibinfo{volume}{102}},
  \bibinfo{pages}{166404} (\bibinfo{year}{2009}).

\bibitem[{\citenamefont{Wang and Upmanyu}(2012)}]{HWang}
\bibinfo{author}{\bibfnamefont{H.}~\bibnamefont{Wang}} and Moneesh Upmanyu,
  \bibinfo{journal}{Phys. Rev. B} \textbf{\bibinfo{volume}{86}},
  \bibinfo{pages}{205411} (\bibinfo{year}{2012}).

\bibitem[{\citenamefont{Sloan et~al.}(2013)\citenamefont{Sloan, {Pacheco
  Sanjuan}, Wang, Horvath, and Barraza-Lopez}}]{us}
\bibinfo{author}{\bibfnamefont{J.~V.} \bibnamefont{Sloan}}, A.~A. Pacheco Sanjuan, Z. Wang, C. Horvath and S. Barraza-Lopez,
  \bibinfo{journal}{Phys. Rev. B} \textbf{\bibinfo{volume}{87}},
  \bibinfo{pages}{155436} (\bibinfo{year}{2013}).

\bibitem[{\citenamefont{Barraza-Lopez et~al.}(2013)\citenamefont{Barraza-Lopez,
  {Pacheco Sanjuan}, Wang, and Vanevi{\'c}}}]{usSSC}
\bibinfo{author}{\bibfnamefont{S.}~\bibnamefont{Barraza-Lopez}}, A.~A. Pacheco Sanjuan, Z. Wang and Mihajlo Vanevi{\'c},
  \bibinfo{journal}{Solid State Comm.} \textbf{\bibinfo{volume}{166}},
  \bibinfo{pages}{70} (\bibinfo{year}{2013}).

\bibitem[{\citenamefont{Vandeparre et~al.}(2011)\citenamefont{Vandeparre,
  Pi{\~n}eirua, Brau, Roman, Bico, Gay, Bao, Lau, Reis, and Damman}}]{newprl}
\bibinfo{author}{\bibfnamefont{H.}~\bibnamefont{Vandeparre}}, M. Pi{\~n}eirua, F. Brau, B. Roman,
J. Bico, C. Gay, W. Bao, C.~N. Lau, P.~M. Reis and Pascal Damman,
  \bibinfo{journal}{Phys. Rev. Lett.} \textbf{\bibinfo{volume}{106}},
  \bibinfo{pages}{224301} (\bibinfo{year}{2011}).

\bibitem[{\citenamefont{Fasolino et~al.}(2007)\citenamefont{Fasolino, Los, and
  Katsnelson}}]{Fasolino1}
\bibinfo{author}{\bibfnamefont{A.}~\bibnamefont{Fasolino}}, J.~H. Los and M.~I. Katsnelson,
  \bibinfo{journal}{Nature Materials} \textbf{\bibinfo{volume}{6}},
  \bibinfo{pages}{858} (\bibinfo{year}{2007}).

\bibitem[{\citenamefont{Zakharchenko et~al.}(2010)\citenamefont{Zakharchenko,
  Rold{\'a}n, Fasolino, and Katsnelson}}]{Zakharenko}
\bibinfo{author}{\bibfnamefont{K.}~\bibnamefont{Zakharchenko}}, R. Rold{\'a}n, A. Fasolino and M.~I. Katsnelson,
  \bibinfo{journal}{Phys. Rev. B} \textbf{\bibinfo{volume}{82}},
  \bibinfo{pages}{125435} (\bibinfo{year}{2010}).

\bibitem[{\citenamefont{Zakharchenko et~al.}(2009)\citenamefont{Zakharchenko,
  Katsnelson, and Fasolino}}]{Katsnelson2}
\bibinfo{author}{\bibfnamefont{K.}~\bibnamefont{Zakharchenko}}, M.~I. Katsnelson and A. Fasolino,
  \bibinfo{journal}{Phys. Rev. Lett.} \textbf{\bibinfo{volume}{102}},
  \bibinfo{pages}{046808} (\bibinfo{year}{2009}).

\bibitem[{\citenamefont{Gibertini et~al.}(2010)\citenamefont{Gibertini,
  Tomadin, Polini, Fasolino, and Katsnelson}}]{PRB2010}
\bibinfo{author}{\bibfnamefont{M.}~\bibnamefont{Gibertini}}, A. Tomadin and M. Polini,
  \bibinfo{journal}{Phys. Rev. B} \textbf{\bibinfo{volume}{81}},
  \bibinfo{pages}{125437} (\bibinfo{year}{2010}).

\bibitem{sanjose}
\bibinfo{author}{\bibfnamefont{P.}~\bibnamefont{San-Jose}}, J. Gonz{\'a}lez and F. Guinea,
  \bibinfo{journal}{Phys. Rev. Lett.} \textbf{\bibinfo{volume}{106}},
  \bibinfo{pages}{045502} (\bibinfo{year}{2011}).


\bibitem[{\citenamefont{Lee et~al.}(2008)\citenamefont{Lee, Wei, Kysar, and
  Hone}}]{Hone1}
\bibinfo{author}{\bibfnamefont{C.}~\bibnamefont{Lee}}, X. Wei, J.~W. Kysar and J. Hone,
  \bibinfo{journal}{Science} \textbf{\bibinfo{volume}{321}},
  \bibinfo{pages}{385} (\bibinfo{year}{2008});
\bibinfo{author}{\bibfnamefont{G.~H.}~\bibnamefont{Lee}}, R.~C. Cooper, S.~J. An, S. Lee, A. van der Zande, N. Petrone, A.~G. Hammerberg,
 C. Lee, B. Crawford, W. Oliver, J.~W. Kysar and J. Hone, \bibinfo{journal}{Science}
  \textbf{\bibinfo{volume}{340}}, \bibinfo{pages}{1073} (\bibinfo{year}{2013}).

\bibitem[{\citenamefont{Suzuura and Ando}(2002)}]{Ando2002}
\bibinfo{author}{\bibfnamefont{H.}~\bibnamefont{Suzuura}} and T. Ando,
  \bibinfo{journal}{Phys. Rev. B} \textbf{\bibinfo{volume}{65}},
  \bibinfo{pages}{235412} (\bibinfo{year}{2002}).

\bibitem[{\citenamefont{Pereira and {Castro-Neto}}(2009)}]{Pereira1}
\bibinfo{author}{\bibfnamefont{V.~M.} \bibnamefont{Pereira}} and A.~H. Castro Neto,
  \bibinfo{journal}{Phys. Rev. Lett.} \textbf{\bibinfo{volume}{103}},
  \bibinfo{pages}{046801} (\bibinfo{year}{2009}).

\bibitem[{\citenamefont{Guinea et~al.}(2010)\citenamefont{Guinea, Katsnelson,
  and Geim}}]{GuineaNatPhys2010}
\bibinfo{author}{\bibfnamefont{F.}~\bibnamefont{Guinea}}, and M.~I. Katsnelson and A.~K. Geim,
  \bibinfo{journal}{Nature Physics} \textbf{\bibinfo{volume}{6}},
  \bibinfo{pages}{30} (\bibinfo{year}{2010}).

\bibitem[{\citenamefont{Vozmediano et~al.}(2010)\citenamefont{Vozmediano,
  Katsnelson, and Guinea}}]{Vozmediano}
\bibinfo{author}{\bibfnamefont{M.~A.~H.} \bibnamefont{Vozmediano}}, and M.~I. Katsnelson, and F. Guinea,
  \bibinfo{journal}{Phys. Rep.} \textbf{\bibinfo{volume}{496}},
  \bibinfo{pages}{109} (\bibinfo{year}{2010}).

\bibitem[{\citenamefont{{de~Juan} et~al.}(2007)\citenamefont{{de~Juan},
  Cortijo, and Vozmediano}}]{deJuanPRB}
\bibinfo{author}{\bibfnamefont{F.}~\bibnamefont{{de~Juan}}}, A. Cortijo and M.~A.~H. Vozmediano,
  \bibinfo{journal}{Phys. Rev. B} \textbf{\bibinfo{volume}{76}},
  \bibinfo{pages}{165409} (\bibinfo{year}{2007}).

\bibitem[{\citenamefont{{de~Juan} et~al.}(2011)\citenamefont{{de~Juan},
  Cortijo, Vozmediano, and Cano}}]{Dejuan2011}
\bibinfo{author}{\bibfnamefont{F.}~\bibnamefont{{de~Juan}}}, A. Cortijo, M.~A.~H. Vozmediano and A. Cano,
  \bibinfo{journal}{Nature Phys.} \textbf{\bibinfo{volume}{7}},
  \bibinfo{pages}{810} (\bibinfo{year}{2011}).

\bibitem[{\citenamefont{{de~Juan} et~al.}(2012)\citenamefont{{de~Juan}, Sturla,
  and Vozmediano}}]{deJuanPRL2012}
\bibinfo{author}{\bibfnamefont{F.}~\bibnamefont{{de~Juan}}}, M. Sturla and M.~A.~H. Vozmediano,
  \bibinfo{journal}{Phys. Rev. Lett.} \textbf{\bibinfo{volume}{108}},
  \bibinfo{pages}{227205} (\bibinfo{year}{2012}).

\bibitem[{\citenamefont{{Neek-Amal} et~al.}(2012)\citenamefont{{Neek-Amal},
  Covaci, and Peeters}}]{Peeters3}
\bibinfo{author}{\bibfnamefont{M.}~\bibnamefont{{Neek-Amal}}}, L. Covaci, and F.~M. Peeters,
  \bibinfo{journal}{Phys. Rev. B} \textbf{\bibinfo{volume}{86}},
  \bibinfo{pages}{041405(R)} (\bibinfo{year}{2012}).

\bibitem[{\citenamefont{{M. R.}~Masir and Peeters}(2013)}]{Peeters4}
\bibinfo{author}{\bibfnamefont{D.~M.} \bibnamefont{{M. R.}~Masir}}, D. Moldovan and F.~M. Peeters,
\bibinfo{journal}{Solid State Comm.} \textbf{\bibinfo{volume}{175-176}},
  \bibinfo{pages}{76}
  (\bibinfo{year}{2013}).

\bibitem[{\citenamefont{Kerner et~al.}(2012)\citenamefont{Kerner, Naumis, and
  G{\'o}mez-Arias}}]{Naumis1}
\bibinfo{author}{\bibfnamefont{R.}~\bibnamefont{Kerner}}, {G.~G.} Naumis and {W.~A.} G{\'o}mez-Arias,
  \bibinfo{journal}{Physica B} \textbf{\bibinfo{volume}{407}},
  \bibinfo{pages}{2002} (\bibinfo{year}{2012}).

\bibitem[{\citenamefont{Ma{\~n}es et~al.}(2013)\citenamefont{Ma{\~n}es, {de
  Juan}, Sturla, and Vozmediano}}]{Manes}
\bibinfo{author}{\bibfnamefont{J.}~\bibnamefont{Ma{\~n}es}}, F. de Juan, M. Sturla and M.~A.~H. Vozmediano,
  \bibinfo{journal}{Phys. Rev. B} \textbf{\bibinfo{volume}{88}},
  \bibinfo{pages}{155405} (\bibinfo{year}{2013}).

\bibitem[{\citenamefont{Levy et~al.}(2010)\citenamefont{Levy, Burke, Meaker,
  Panlasigui, Zettl, Guinea, Castro-Neto, and Crommie}}]{Crommie}
\bibinfo{author}{\bibfnamefont{N.}~\bibnamefont{Levy}}, S.~A. Burke, K.~L. Meaker, M. Panlasigui, A. Zettl, F. Guinea,
 A.~H. Castro-Neto and M.~F. Crommie,
 \bibinfo{journal}{Science}
  \textbf{\bibinfo{volume}{329}}, \bibinfo{pages}{544} (\bibinfo{year}{2010}).

\bibitem[{\citenamefont{Xu et~al.}(2012)\citenamefont{Xu, Yang, Barber,
  Ackerman, Schoelz, Qi, Kornev, Dong, Bellaiche, Barraza-Lopez
  et~al.}}]{usold}
\bibinfo{author}{\bibfnamefont{P.}~\bibnamefont{Xu}}, Y. Yang, S.~D. Barber, M.~L. Ackerman, J.~K. Schoelz, D. Qi,
I.~A. Kornev, L. Dong, L. Bellaiche, S. Barraza-Lopez and P.~M. Thibado,
\bibinfo{journal}{Phys. Rev. B}
\textbf{\bibinfo{volume}{85}}, \bibinfo{pages}{121406(R)}
(\bibinfo{year}{2012});
\bibinfo{author}{\bibfnamefont{R.}~\bibnamefont{Zan}}, C. Muryn, U. Bangert, P. Mattocks, P. Wincott, D. Vaughan,
 X. Li, L. Colombo, R.~S. Ruoff, B. Hamilton and K.~S. Novoselov,
\bibinfo{journal}{Nanoscale}
  \textbf{\bibinfo{volume}{4}}, \bibinfo{pages}{3065} (\bibinfo{year}{2012});
\bibinfo{author}{\bibfnamefont{F.~R.} \bibnamefont{Eder}}, J. Kotakoski, K. Holzweber, C. Mangler, V. Skakalova and J.~C. Meyer,
\bibinfo{journal}{Nano Lett.} \textbf{\bibinfo{volume}{13}},
  \bibinfo{pages}{1934} (\bibinfo{year}{2013}).


\bibitem[{\citenamefont{Klimov et~al.}(2012)\citenamefont{Klimov, Jung, Zhu,
  Li, Wright, Solares, Newell, Zhitenev, and Stroscio}}]{stroscio}
\bibinfo{author}{\bibfnamefont{N.~N.} \bibnamefont{Klimov}}, S. Jung, S. Zhu, T. Li, C.~A. Wright, S.~D. Solares, D.~B. Newell, N.~B. Zhitenev and J.~A. Stroscio,
\bibinfo{journal}{Science}
  \textbf{\bibinfo{volume}{336}}, \bibinfo{pages}{1557} (\bibinfo{year}{2012}).

\bibitem[{\citenamefont{Kitt et~al.}(2013{\natexlab{b}})\citenamefont{Kitt,
  Pereira, Swan, and Goldberg}}]{Kitt2013}
\bibinfo{author}{\bibfnamefont{A.~L.} \bibnamefont{Kitt}}, V.~M. Pereira, A.~K. Swan and B.~B. Goldberg,
  \bibinfo{journal}{Phys. Rev. B} \textbf{\bibinfo{volume}{87}},
  \bibinfo{pages}{159909(E)} (\bibinfo{year}{2013}{\natexlab{b}});
\bibinfo{author}{\bibfnamefont{F.}~\bibnamefont{de~Juan}}, J.~L. Ma{\~n}es and M.~A.~H. Vozmediano,
 \bibinfo{journal}{Phys. Rev. B}
  \textbf{\bibinfo{volume}{87}}, \bibinfo{pages}{165131}
  (\bibinfo{year}{2013}); \bibinfo{author}{\bibfnamefont{M.}~\bibnamefont{Oliva-Leyva}} and G.~G. Naumis,
  \bibinfo{journal}{Phys. Rev. B} \textbf{\bibinfo{volume}{88}},
  \bibinfo{pages}{085430} (\bibinfo{year}{2013}).


\bibitem[{\citenamefont{Lee}(1997)}]{Lee}
\bibinfo{author}{\bibfnamefont{J.}~\bibnamefont{Lee}},
  \emph{\bibinfo{title}{Riemannian Manifolds: An introduction to curvature}}.
  (\bibinfo{publisher}{Springer}, \bibinfo{address}{New York},
  \bibinfo{year}{1997}), \bibinfo{edition}{1st} ed.

\bibitem[{\citenamefont{{do Carmo}}(1976)}]{doCarmo}
\bibinfo{author}{\bibfnamefont{M.}~\bibnamefont{{do Carmo}}},
  \emph{\bibinfo{title}{Differential Geometry of Curves and Surfaces}}
  (\bibinfo{publisher}{Prentice Hall}, \bibinfo{address}{New Jersey},
  \bibinfo{year}{1976}), \bibinfo{edition}{1st} ed.

\bibitem[{\citenamefont{Hollerer and Celigoj}(2012)}]{M4}
\bibinfo{author}{\bibfnamefont{S.}~\bibnamefont{Hollerer}} and C.~C. Celigoj,
  \bibinfo{journal}{Comput. Mech.}
 \textbf{\bibinfo{volume}{51}},
  \bibinfo{pages}{765} (\bibinfo{year}{2013}).



\bibitem[{\citenamefont{Chen et~al.}(2009)\citenamefont{Chen, Rosenblatt,
  Bolotin, Kalb, Kim, Kymissis, Stormer, Heinz, and Hone}}]{ChenNatureNano}
\bibinfo{author}{\bibfnamefont{C.}~\bibnamefont{Chen}}, S. Rosenblatt, K.~I. Bolotin, W. Kalb, P. Kim, I. Kymissis, H.~L. Stormer,
 T.~F. Heinz and J. Hone,
  \bibinfo{journal}{Nature Nano} \textbf{\bibinfo{volume}{4}},
  \bibinfo{pages}{861} (\bibinfo{year}{2009}).

\bibitem[{\citenamefont{Arroyo and Belytschko}(2003)}]{M2}
\bibinfo{author}{\bibfnamefont{M.}~\bibnamefont{Arroyo}} and T. Belytschko,
  \bibinfo{journal}{Phys. Rev. Lett.} \textbf{\bibinfo{volume}{91}},
  \bibinfo{pages}{215505} (\bibinfo{year}{2003}).

\bibitem[{\citenamefont{Zhang et~al.}(2011)\citenamefont{Zhang, Akatyeva, and
  Dumitrica}}]{Dumitrica2011}
\bibinfo{author}{\bibfnamefont{D.-B.} \bibnamefont{Zhang}}, E. Akatyeva and T. Dumitrica,
  \bibinfo{journal}{Phys. Rev. Lett.} \textbf{\bibinfo{volume}{106}},
  \bibinfo{pages}{255503} (\bibinfo{year}{2011}).

\bibitem[{\citenamefont{Kitt et~al.}(2013{\natexlab{a}})\citenamefont{Kitt, Qi,
  R{\'e}mi, Park, Swan, and Goldberg}}]{Kitt2}
\bibinfo{author}{\bibfnamefont{A.~L.} \bibnamefont{Kitt}}, Z. Qi, S. R{\'e}mi, H.~S. Park, A.~K. Swan and B.~B. Goldberg,
  \bibinfo{journal}{Nano Lett.} \textbf{\bibinfo{volume}{13}},
  \bibinfo{pages}{2605} (\bibinfo{year}{2013}).

\bibitem[{\citenamefont{Huertas-Hernando
  et~al.}(2006)\citenamefont{Huertas-Hernando, Guinea, and
  Brataas}}]{Huertas-Hernando}
\bibinfo{author}{\bibfnamefont{D.}~\bibnamefont{Huertas-Hernando}}, F. Guinea and A. Brataas,
  \bibinfo{journal}{Phys. Rev. B} \textbf{\bibinfo{volume}{74}},
  \bibinfo{pages}{155426} (\bibinfo{year}{2006}).


\bibitem{ACSNano}
   \bibinfo{author}{\bibfnamefont{A.~A.} \bibnamefont{{Pacheco~Sanjuan}}}, M. Mehboudi, E.~O. Harriss, H. Terrones and S. Barraza-Lopez,
  \bibinfo{journal}{ACS Nano},  DOI: 10.1021/nn406532z. (\bibinfo{year}{2014}).


\bibitem{pe2013}
   \bibinfo{author}{\bibfnamefont{M.} \bibnamefont{{Neek-Amal}}}, L. Covaci, K. Shakouri, and F.~M. Peeters,
  \bibinfo{journal}{Phys. Rev. B} \textbf{\bibinfo{volume}{88}},
  \bibinfo{pages}{115428} (\bibinfo{year}{2013}).

\bibitem[{\citenamefont{Bobenko et~al.}(2008)\citenamefont{Bobenko,
  Schr{\"o}der, Sullivan, and Ziegler}}]{Math1}
\bibinfo{editor}{\bibfnamefont{A.~I.} \bibnamefont{Bobenko}},
  \bibinfo{editor}{\bibfnamefont{P.}~\bibnamefont{Schr{\"o}der}},
  \bibinfo{editor}{\bibfnamefont{J.~M.} \bibnamefont{Sullivan}},
  \bibnamefont{and} \bibinfo{editor}{\bibfnamefont{G.~M.}
  \bibnamefont{Ziegler}}, eds., \emph{\bibinfo{title}{Discrete Differential
  Geometry}}, vol.~\bibinfo{volume}{38} of \emph{\bibinfo{series}{Oberwolfach
  Seminars}} (\bibinfo{publisher}{Springer}, \bibinfo{address}{Germany},
  \bibinfo{year}{2008}), \bibinfo{edition}{1st} ed.

\bibitem[{\citenamefont{Bobenko and Suris}(2009)}]{Math2}
\bibinfo{editor}{\bibfnamefont{A.~I.} \bibnamefont{Bobenko}} \bibnamefont{and}
  \bibinfo{editor}{\bibfnamefont{Y.~B.} \bibnamefont{Suris}},
  \emph{\bibinfo{title}{Discrete Differential Geometry: Integrable Structure}}. (\bibinfo{publisher}{AMS}, \bibinfo{address}{USA},
  \bibinfo{year}{2009}), \bibinfo{edition}{1st} ed.

\bibitem[{\citenamefont{Xu and Xu}(2009)}]{chinese}
\bibinfo{author}{\bibfnamefont{Z.}~\bibnamefont{Xu}} and G. Xu, \bibinfo{journal}{Comp.
  Math. Appl.} \textbf{\bibinfo{volume}{57}}, \bibinfo{pages}{1187}
  (\bibinfo{year}{2009}).


\bibitem{SI}
See supplementary information.

\bibitem[{\citenamefont{Gomes et~al.}(2012)\citenamefont{Gomes, Mar, Ko,
  Guinea, and Manoharan}}]{MolecularGraphene}
\bibinfo{author}{\bibfnamefont{K.~K.} \bibnamefont{Gomes}}, W. Mar, W. Ko, F. Guinea and H.~C. Manoharan,
  \bibinfo{journal}{Nature} \textbf{\bibinfo{volume}{483}},
  \bibinfo{pages}{306} (\bibinfo{year}{2012}).

\bibitem[{\citenamefont{Cirak et~al.}(2000)\citenamefont{Cirak, Ortiz, and
  Schr{\"o}der}}]{M1}
\bibinfo{author}{\bibfnamefont{F.}~\bibnamefont{Cirak}}, M. Ortiz and P. Schr{\"o}der,
  \bibinfo{journal}{Int. J. Numer. Meth. Engng.} \textbf{\bibinfo{volume}{47}},
  \bibinfo{pages}{2039} (\bibinfo{year}{2000}).

\bibitem[{\citenamefont{Ando}(2000)}]{Ando2000}
\bibinfo{author}{\bibfnamefont{T.}~\bibnamefont{Ando}}, \bibinfo{journal}{J.
  Phys. Soc. Jp.} \textbf{\bibinfo{volume}{69}}, \bibinfo{pages}{1757}
  (\bibinfo{year}{2000}).

\bibitem{Katsnelson}
\bibinfo{author}{\bibfnamefont{M.~I.} \bibnamefont{Katsnelson}}.
Graphene: Carbon in two dimensions. Cambridge U. Press (2012).

\bibitem[{\citenamefont{Choi et~al.}(2010)\citenamefont{Choi, Jhi, and
  Son}}]{YWSon}
\bibinfo{author}{\bibfnamefont{S.-M.} \bibnamefont{Choi}}, S.-H. Jhi and Y.-W. Son,
  \bibinfo{journal}{Phys. Rev. B} \textbf{\bibinfo{volume}{81}},
  \bibinfo{pages}{081407} (\bibinfo{year}{2010}).

\bibitem[{\citenamefont{Lukose et~al.}(2007)\citenamefont{Lukose, Shankar, and
  Baskaran}}]{Lukose}
\bibinfo{author}{\bibfnamefont{V.}~\bibnamefont{Lukose}}, R. Shankar and G. Baskaran,
  \bibinfo{journal}{Phys. Rev. Lett.} \textbf{\bibinfo{volume}{98}},
  \bibinfo{pages}{116802} (\bibinfo{year}{2007}).



\bibitem[{\citenamefont{Kim et~al.}(2011)\citenamefont{Kim, Blanter, and
  Ahn}}]{Blanter}
\bibinfo{author}{\bibfnamefont{K.-J.} \bibnamefont{Kim}}, Ya.~M. Blanter and K.-H. Ahn,
  \bibinfo{journal}{Phys. Rev. B} \textbf{\bibinfo{volume}{84}},
  \bibinfo{pages}{081401(R)} (\bibinfo{year}{2011}); \bibinfo{author}{\bibfnamefont{G.~M.~M.}~\bibnamefont{Wakker}}, R.~P. Tiwari and M. Blaauboer,
  \bibinfo{journal}{Phys. Rev. B} \textbf{\bibinfo{volume}{84}},
  \bibinfo{pages}{195427} (\bibinfo{year}{2011}).
\end{thebibliography}
\end{document}